\documentclass[structabstract]{aa}
\usepackage{graphicx}
\usepackage{txfonts}
\usepackage{natbib}
\bibpunct{(}{)}{;}{a}{}{,} 
\usepackage{lscape}
\usepackage{aalongtable}
\begin{document}

   \title{Accretion-related properties of Herbig Ae/Be stars}
\subtitle{Comparison with T Tauris}

   \author{I. Mendigut\'{\i}a\inst{1}
\and
A. Mora\inst{2}
\and         
B. Montesinos\inst{3}         
\and
C. Eiroa\inst{1}
\and
G. Meeus\inst{1}
\and
B. Mer\'{\i}n\inst{4}         
\and
R.D. Oudmaijer\inst{5}
}

   \offprints{Ignacio Mendigut\'{\i}a\\
              \email{Ignacio.Mendigutia@cab.inta-csic.es}}

   \institute{$^{1}$Departamento de F\'{\i}sica Te\'{o}rica, M\'{o}dulo 15,
     Facultad de Ciencias, Universidad Aut\'{o}noma de Madrid, PO Box
     28049, Cantoblanco, Madrid, Spain.\\
     $^{2}$GAIA Science Operations Centre, ESA, European Space Astronomy Centre, PO Box 78, 28691, Villanueva de la Ca\~nada, Madrid,
     Spain.\\
     $^{3}$Departamento de Astrof\'{\i}sica, Centro de Astrobiolog\'{\i}a (INTA-CSIC), ESAC Campus, P.O. Box 78, 
     28691 Villanueva de la Ca\~nada, Madrid, Spain.\\
     $^{4}$Herschel Science Centre, ESA, European Space Astronomy Centre, P.O. Box 78, 28691, Villanueva de la Ca\~nada, Madrid,
     Spain.\\
     $^{5}$School of Physics \& Astronomy, University of Leeds, Woodhouse
     Lane, Leeds LS2 9JT, UK.\\}

   \date{Received ; accepted}

 
  \abstract{The mass accretion rate ($\dot{M}_{\rm acc}$) is a key parameter that has not accurately been determined for a wide sample of Herbig Ae/Be (HAeBe) stars until recently.}
  {We look for trends relating $\dot{M}_{\rm acc}$ and the stellar ages ($t$), spectral energy distributions (SEDs), and disk masses for a sample of 38 HAeBe stars, comparing them to analogous correlations found for classical T Tauri stars. Our goal is ultimately to shed light on the timescale and physical processes that drive evolution of intermediate-mass pre-main sequence objects.}
  {Mass accretion rates obtained by us in a previous work were related to several stellar and disk parameters: the age of the stars was compiled from the literature, SEDs were classified according to their shape and the wavelength where the IR excess starts, near- and mid-IR colour excesses were computed, and disk masses were estimated from mm fluxes.}
  {$\dot{M}_{\rm acc}$ decreases with the stellar age, showing a dissipation timescale $\tau$ = 1.3$^{+1.0}_{-0.5}$ Myr from an exponential law fit, while a power law yields $\dot{M}_{\rm acc}$(t) $\propto$ t$^{-\eta}$, with $\eta$ = 1.8$^{+1.4}_{-0.7}$. This result is based on our whole HAeBe sample (1 -- 6 M$_\odot$), but the accretion rate decline most probably depends on smaller stellar mass bins. The near-IR excess is higher and starts at shorter wavelengths ($J$ and $H$ bands) for the strongest accretors. Active and passive disks are roughly divided by $\sim$ 2 $\times$ 10$^{-7}$ M$_{\sun}$ yr$^{-1}$. The mid-IR excess and the SED shape from the Meeus et al. classification are not correlated with $\dot{M}_{\rm acc}$. Concerning disk masses, we find  $\dot{M}_{\rm acc}$ $\propto$ M$_{disk}$$^{1.1 \pm 0.3}$. Most stars in our sample with signs of inner dust dissipation typically show accretion rates ten times lower and disk masses three times smaller than the remaining objects.}
  {The trends relating $\dot{M}_{\rm acc}$ with the near-IR excess and M$_{disk}$ extend those found for T Tauri stars, and are consistent with viscous disk models. The differences in the inner gas dissipation timescale, and the relative position of the stars with signs of inner dust clearing in the $\dot{M}_{\rm acc}$--M$_{disk}$ plane, could be suggesting a slightly faster evolution, and that a different process - such as photoevaporation - plays a more relevant role in dissipating disks in the HAeBe regime compared to T Tauri stars. Our conclusions must consider the mismatch between the disk mass estimates from mm fluxes and the disk mass estimates from accretion, which we also find in HAeBe stars.}


   \maketitle
%

\section{Introduction}
\label{Section:introduction}
Nowadays there is consensus that practically all stars are formed from accretion disks \citep{Kraus10}, even if these have sometimes very short dissipation timescales. The mass accretion rate ($\dot{M}_{\rm acc}$) is a major parameter that drives the star-disk evolution, given that a significant fraction of the stellar mass acquired during the pre-main sequence (PMS) phase is accreted from the disk. Circumstellar disks are also dissipated through other physical processes, including photo-evaporation, dust settling, and dynamical interactions with forming planets \citep[see e.g.][]{Cieza08,WilliamsCieza11}. Their relative influence could in turn be estimated from the amount of gas accreted by the star \citep{Alexander07,Najita07}.

Mass accretion rates in classical T Tauri (CTT) stars are mainly derived from the UV emission excess and spectroscopic veiling caused by the accretion shocks, as well as from magnetospheric accretion line modelling and empirical correlations with the luminosity of several emission lines \citep[see the review by][and references therein]{Calvet00}. The CTTs show a decrease in $\dot{M}_{\rm acc}$ of about one or two orders of magnitude in the first Myr \citep[e.g.][]{Fang09}, which gives an estimate of the inner gas dissipation timescale. This is intimately related to the inner dust disk lifetime \citep{Hartigan95}, indicating that both gas and dust close to the star vanish at a similar rate. Measurements of dust emission at longer wavelengths provide estimates of the disk mass (M$_{disk}$) and point to an inside-out dissipation of the circumstellar environment \citep[][and references therein]{Cieza08}. The so-called transitional disks,  where the IR emission is weak or negligible up to $\lambda$ $\gtrsim$ 10 $\mu$m, are particularly interesting. Stars with this type of spectral energy distribution (SED) are interpreted to be in a fast evolutionary stage that bridges the gap between the optically thick disk phase and the optically thin debris disk one \citep{Wolk96,Cieza07}. An analysis of the M$_{disk}$--$\dot{M}_{\rm acc}$ plane can be useful to discern between the different physical processes that drive disk dissipation. In particular, the region occupied by several CTTs with transitional disks in that plane is compatible with the formation of Jovian planets \citep{Najita07}. 

Herbig Ae/Be (HAeBe) objects \citep{Waters06} are the intermediate-mass ($\sim$ 1--10 M$_{\sun}$) counterparts of CTT stars. The HAeBes are the most massive objects to experience an optically visible pre-main sequence phase, bridging the transition between low- and high-mass stars. Although these are highly interesting objects - e.g., planet formation could be more efficient in stars more massive than the Sun \citep{KennedyKenyon08,Bowler10,Boss11} - our knowledge about the properties of the HAeBe regime is much more limited than that for the T Tauri stars. The main reason is the comparatively smaller HAeBe sample, caused by the faster evolution of massive stars to the main-sequence, and by the fact that star formation favours lower mass objects, as the shape of the initial-mass function suggests \citep{Salpeter95}. 

The optical photometry in \citet{Oudmaijer01} and the simultaneous optical spectroscopy in \citet[][Paper I hereafter]{Mendigutia11a} were recently used to derive accretion rate estimates and empirical relations with emission lines for 38 HAeBe stars \citep[][Paper II hereafter]{Mendigutia11b}. The accretion rates in Paper II are the first obtained for a wide sample of HAeBe stars from a specific modelling of the UV Balmer excess and constitute the most reliable estimates for this regime to date. The present work closes this series of papers. Here, we analyse possible relationships between the mass accretion rates and several stellar and disk parameters of the 38 HAeBes studied in Papers I and II, and compare them to the better known properties for CTTs. In particular, we look for trends with the stellar age, properties of the SED, and disk mass. The ultimate objective is to contribute to the knowledge of circumstellar disk dissipation in HAeBe stars. The paper is organized as follows. Section \ref{Section: Sample and data} describes some general properties of the sample. The SED properties and disk masses are treated separately in Sects. \ref{Section:SEDs} and \ref{Section:Mdisk}, respectively. The correlations with the mass accretion rate are presented and analysed in Sect. \ref{Section:analysis}. The discussion is included in Sect. \ref{Section:discussion}, and the summary and conclusions in Sect. \ref{Section:conclusion}.

\section{Sample properties and data}
\label{Section: Sample and data}

The sample consists of 38 HAeBe stars covering almost all such objects in the northern hemisphere from the catalogue of \citet{The94} (see Papers I and II). All stars show IR excess and variable emission lines \citep[][Paper I]{Merin04}. Columns 2 to 8 in Table \ref{Table:sample} include the effective temperatures, distances, ages, mass accretion rates, and information on possible stellar companions. The parameters in Cols. 9 to 16 are described in the following sections. The spectral types cover the Ae and late-type Be regime, including ten intermediate-mass F and G stars \citep{Mora01}. The stellar age covers the pre-main sequence phase for our sample, up to $\sim$ 15 Myr. Most ages were determined by \citet{Montesinos09}, with a typical uncertainty of 35$\%$. The range in stellar mass and mass accretion rate is $\sim$ 1--6 M$_{\odot}$ and 10$^{-9}$--10$^{-5}$ M$_{\odot}$ yr$^{-1}$, with a typical -median- value of 2.4 $\times$ 10$^{-7}$ M$_{\odot}$ yr$^{-1}$. The mass accretion rates are those reproducing the observed Balmer excesses from a magnetospheric accretion shock model (Paper II). The accretion rates for \object{R Mon}, \object{VY Mon}, \object{VV Ser}, and \object{LkHa 234} were estimated extrapolating the calibrations with the emission line luminosities provided in that work. Their large uncertainties reflect the difficulty to model the strong Balmer excess of these stars. In case of multiplicity, most of the stellar and accretion parameters are in principle referred to the brightest component, but some contamination cannot be excluded for objects with close companions (d $<$ 1''; see references in Table \ref{Table:sample}). The high percentage of binaries is typical for HAeBe stars \citep{Wheelwright10}.

\begin{table*}
\renewcommand\tabcolsep{2pt}
\centering
\caption{Sample of stars.}
\label{Table:sample}
\begin{tabular}{llllllllcrcrrccr}
\hline\hline
Star & T & d & Age & $\dot{M}_{\rm acc}$ & d(SpT') & $\delta$$K$ &
Ref&M01 & IR$_{start}$ & E($H-K$)$_0$ & [12]-[25] & [25]-[60]& T$_D$ & $\beta$ & M$_{disk}$ \\ 
 & (K)  & (pc)& (Myr) & [M$_{\sun}$ yr$^{-1}$] &('')& (mag) &  & &(band)	 &  (mag)  & (mag) & (mag)  & (K)  & &[M$_{\odot}$]   \\
\hline
[1] & [2] & [3] & [4] & [5] & [6] & [7] & [8] & [9] & [10] & [11] & [12] & [13] & [14] & [15] & [16]\\
\hline
\object{HD 31648} & 8250   & 146     & 6.7	       &$<$-7.23       & ...	      & ...	 & ...  	 &II&H   &0.71&-0.18&0.08  &21 &0.9&-1.27$\pm$0.03\\
\object{HD 34282} & 9550$^A$   & 164$^A$ & $>$ 7.8$^{A}$  &$<$-8.30       & ...	      & ...	 & ...  	 &I &K   &0.57&0.92 &2.01  &38 &1.4&-1.94$\pm$0.04\\
\object{HD 34700} & 6000$^B$   & 336$^E$ & 3.4$^{B}$      &$<$-8.30       & 5.2(M1-2)    &3.3	 & (1)  	 &I &$>$K&0.18&2.17 &1.26  &16 &1.5&-1.86$\pm$0.04\\
	 &	    &	      & 	       &	       & 9.2(M2-3)    &4.2  & (1)		 &&&&&&&&\\
\object{HD 58647} & 10500  & 543     & 0.4	       &-4.84$\pm$0.22 & $>$0.5(?)    & ?	& (2)		 &II&H   &0.65&-0.59&-1.96 &...&...&... 	  \\
\object{HD 141569}& 9550$^A$ & 99$^A$  & 6.7$^{A}$      &-6.89$\pm$0.40 & 7.6(M2)      & 1.8$^{*1}$& (3) 	 &II&$>$K&0.04&1.33 &1.18  &33 &1.2&-3.41$\pm$0.19\\
         &	    &	      & 	       &	       & 9.0(M4)      & 2.4$^{*1}$& (4) 	 &&&&&&&&\\
\object{HD 142666}& 7590$^A$    & 145$^A$ & 5.1$^{A}$      &-6.73$\pm$0.26 & ...	     & ...	& ...		 &II&K   &0.56&0.29 &-0.48 &21 &0.6&-1.73$\pm$0.03\\
\object{HD 144432}& 7410$^A$  & 145$^A$ & 5.3$^{A}$      &$<$-7.22       & 1.4(K5Ve)    & 2.4	& (4,5) 	 &II&H   &0.61&0.24 &-0.53 &16 &1.0&-2.15$\pm$0.04\\
\object{HD 150193}& 8970   & 203     & 5.0	       &-6.12$\pm$0.14 & 1.1(F9Ve)    & 2.2	& (4,6) 	 &II&J   &0.67&0.03 &-0.87 &26 &0.2&-1.98$\pm$0.04\\
\object{HD 163296}& 9250  & 130     & 5.0	       &$<$-7.52       & ...	     & ...	& ...		 &II&H   &0.76&0.15 &0.32  &29 &0.5&-1.18$\pm$0.02\\
\object{HD 179218}& 9500   & 201     & 3.3	       &$<$-7.30       & 2.5(?)       & 6.6	& (7,8) 	 &I &K   &0.64&0.67 &-0.41 &29 &1.3&-1.84$\pm$0.04\\
\object{HD 190073}& 9500  & 767     & 0.6	       &-5.00$\pm$0.25 & ...	     & ...	& ...		 &II&H   &0.74&-0.28&-1.15 &...&...&... 	  \\
\object{AS 442}   & 11000    & 826     & 1.5	       &-5.08$\pm$0.11 &?(?)	     & ?	& (9)		 &I &J   &0.86&-0.71&1.38  &...&...&... 	  \\
\object{VX Cas}   & 10000   & 619     & 6.4	       &-6.44$\pm$0.22 & 5.3(?)       & 4.8	& (8)	 &II&K   &0.91&0.59 &-0.69 &28 &1.0&$<$-1.92	  \\
\object{BH Cep}   &6460$^A$  & 450$^A$ & 8.2$^{A}$      &$<$-8.30       &...	     &...	&...	 &II&K   &0.55&-1.53&0.19  &...&...&... 	  \\
\object{BO Cep}   &6610$^A$     & 400$^A$ & 11.2$^{A}$     &-6.93$\pm$0.28 &...	     &...	&...		 &I &$>K$&0.19&1.52 &1.75  &...&...&... 	  \\
\object{SV Cep}   & 10250   & 596     & 5.2	       &-6.30$\pm$0.20 & ...	     & ...	& ...	 &II&H   &0.72&0.23 &-0.73 &24 &1.0&-1.82$\pm$0.12\\
\object{V1686 Cyg}& 6170$^A$    & 980$^A$ & $<$ 0.2$^{A}$  &-5.23$\pm$0.41 & ...	     & ...	& ...		 &I &K   &1.05&0.53 &1.50  &16 &1.0&$<$-0.42	  \\
\object{R Mon}    & 12020$^A$ & 800$^A$ & $<$ 0.01$^{A}$ &(-4$\pm$1)     & 0.7(K1V)     & 6.0$^{*2}$& (10)	  &I &H   &1.49&0.96 &-0.09 &56 &0.7&-0.78$\pm$0.04\\
\object{VY Mon}   &12020$^A$    & 800$^A$ & $<$ 0.01$^{A}$ &(-4$\pm$1)     &...	     &...	&...	 &I &J   &0.97&0.67 &0.57  &32 &2.4&-0.45$\pm$0.04\\
\object{51 Oph}   & 10250& 142     & 0.7	       &$<$-7.00       & ...	     & ...	& ...		 &II&$>$K&0.42&-0.47&-2.46 &...&...&... 	  \\
\object{KK Oph}   & 7590$^A$   & 160$^A$ & 3.9$^{A}$      &-5.85$\pm$0.15 & 1.5(G6Ve)    & 2.5	& (4,6)  &II&K   &1.33&-0.03&-0.48 &16 &0.1&-1.91$\pm$0.04\\
\object{T Ori}    & 9750  & 472     & 4.0	       &-6.58$\pm$0.40 & 7.7	     & $>$4.5	& (11)  	 &I &$>$K&0.86&...  &...   &30 &1.0&-0.85$\pm$0.09\\
\object{BF Ori}   & 8970  & 603     & 3.2	       &$<$-8.00       & ...	     & ...	& ...	 &II&K   &0.60&-0.20&1.22  &24 &1.0&-1.88$\pm$0.14\\
\object{CO Ori}   & 6310$^A$   & 450$^A$ & $<$ 0.1$^{A}$  &-5.20$\pm$0.18 & 2.0(?)       & 3.2$^{*3}$& (12)	  &II&H   &0.65&-0.13&0.40  &15 &1.0&$<$-2.11	  \\
\object{HK Ori}   & 8510$^A$    & 460$^A$ & 1.0$^{A}$      &-5.24$\pm$0.12 & 0.4(K3)      & 2.30	& (11,13)&II&$>$K&0.86&0.08 &-0.99 &20 &1.0&$<$-0.81	  \\
\object{NV Ori}   & 6750$^C$ & 450$^F$ & 4.4$^{C}$      &$<$-8.30       & ...	     & ...	& ...		 &II&K   &0.51&...  &...   &...&...&... 	  \\
\object{RY Ori}   & 6310$^A$   & 460$^A$ & 1.8$^{A}$      &-6.65$\pm$0.33 & ...	     & ...	& ...	 &II&H   &0.49&-0.06&1.27  &...&...&... 	  \\
\object{UX Ori}   & 8460   & 517     & 4.5	       &$<$-6.77       &$\geq$ 0.02(?)& ?	& (12)   &II&H   &0.72&0.35 &-0.28 &28 &0.6&-1.49$\pm$0.04\\
\object{V346 Ori} & 9750   & 586     & 3.5	       &-5.99$\pm$0.17 & ...	     & ...	& ...		 &I &K   &0.47&1.51 &1.63  &...&...&... 	  \\
\object{V350 Ori} & 8970   & 735     & 5.5	       &-6.66$\pm$0.24 & 0.3(?)       & 3.2	& (8)		 &II&$>$K&0.71&0.67 &-0.42 &...&...&... 	  \\
\object{XY Per}   & 9750    & 347     & 2.5	       &-5.86$\pm$0.20 & 1.2(?)       & 0.0	& (6)		 &II&H   &0.73&0.06 &0.21  &...&...&... 	  \\
\object{VV Ser}   & 13800  & 614     & 1.2	       &(-5.2$\pm$0.8) & ...	     & ...	& ...	 &II&H   &0.93&-0.22&0.56  &45 &1.0&$<$-1.61	  \\
\object{CQ Tau}   & 6800$^{B}$   & 130$^G$ & 7.7$^{B}$      &$<$-8.30       & ...	     & ...	& ...	 &I &K   &0.73&1.31 &0.06  &14 &0.6&-1.60$\pm$0.03\\
\object{RR Tau}   & 10000  & 2103    & 0.4	       &-4.11$\pm$0.16 & ?(?)	     & ?	& (7)	 &II&H   &0.88&0.25 &0.77  &28 &1.0&$<$-0.33	  \\
\object{RY Tau}   & 5770$^C$ & 134$^F$ & 6.5$^D$        &-7.65$\pm$0.17 & ...	     & ...	& ...		 &II&H   &0.68&0.43 &-0.58 &14 &1.0&-1.50$\pm$0.04\\
\object{PX Vul}   & 6760$^A$    & 420$^A$ & 14$^{A}$       &-6.72$\pm$0.16 & ...	     & ...	& ...		 &II&K   &0.57&0.49 &-0.68 &...&...&... 	  \\
\object{WW Vul}   & 8970   & 696     & 3.7	       &-6.38$\pm$0.70 & ...	     & ...	& ...	 &II&H   &0.83&0.31 &-0.38 &24 &1.0&-1.51$\pm$0.04\\
\object{LkHa 234} & 12900$^A$   & 1250$^A$& $<$ 0.01$^{A}$ &(-4.1$\pm$0.4) & 2.7(?)       & $>$ 5.20 & (11)   &I &H   &0.88&1.82 &2.35  &65 &2.5&0.31$\pm$0.04 \\
\hline
\end{tabular}

\begin{minipage}{18.5cm}

  \textbf{Notes.} Columns 1 to 5 list the name of the star, stellar effective temperature, distance, age, and mass accretion rate. Unless otherwise indicated, the stellar temperatures, distances and ages are taken from \citet{Montesinos09}. Accretion rates (on log scale) are taken from Paper II. Columns 6 to 8 refer to possible stellar companions. Column 6 lists their angular distance and spectral type, Col. 7 their difference in $K$ magnitude ($K$$_s$ for $^{*1}$, $K'$ for $^{*2}$ and Hipparcos H$_p$ for $^{*3}$), and Col. 8 the corresponding references. \object{HD 34700} is a triple visual system, the brightest object of which is a spectroscopic binary (1). \object{HD 141569} is a triple system (3). Columns 9 to 13 list the SED classification from the \citet{Meeus01} scheme, the photometric band according to the shortest wavelength where the IR excess is apparent, the intrinsic near-IR colour excess, and the IRAS colours. Columns 14 to 16 list the dust temperature, $\beta$ value, and disk mass (on log scale). These were obtained from the fluxes at 1.3 mm, except for \object{HD 34700}, \object{V1686 Cyg}, \object{RY Tau} and \object{LkHa 234} (see Table \ref{Table:photometrylong}).\\
  \textbf{References.} $^A$\citet{Manoj06}, $^B$\citet{AlonsoAlbi09} and references therein, $^C$\citet{Merin04}, $^D$\citet{Siess99} $^E$\citet{Acke05}, $^F$\citet{Blondel06}, $^G$\citet{GarciaLopez06}. (1): \citet{Sterzik05}, (2): \citet{Baines06}, (3): \citet{Weinberger00}, (4): \citet{Carmona07}, (5): \citet{Perez04}, (6): \citet{Pirzkal97}, (7): \citet{Wheelwright10}, (8): \citet{Thomas07}, (9): \citet{Corporon99}, (10): \citet{Close97}, (11): \citet{Leinert97}, (12): \citet{Bertout99}, (13): \citet{Smith05}.
\end{minipage}
\end{table*}

\subsection{Spectral energy distributions}
\label{Section:SEDs}

Multi-wavelength photometry compiled from the literature is listed in Tables \ref{Table:photometry}, \ref{Table:photometrylong} and \ref{Table:photometryfar}. Figure \ref{Figure:SEDs} shows the SEDs of the stars in the sample. The best fits obtained using the online SED fitting tool from the 2-D radiative transfer models in \citet{Robitaille06,Robitaille07}\footnote{http://caravan.astro.wisc.edu/protostars/} are overplotted. The use of simultaneous $UBVRIJHK$ photometry selected at the brightest $V$ state from the data in \citet{Oudmaijer01} and \citet{Eiroa02} allow us to minimize the influence of obscuring effects from circumstellar material, providing the best possible photospheric-inner disk fitting. The wavelength where the IR excess becomes apparent can thus be derived for each object (Sect. \ref{Section:irexcess}). Photometry at wavelengths longer than 160 $\mu$m was not used if its inclusion made the stellar and inner-mid disk fitting worse - note that Robitaille's models were originally optimized for the 1 to 100 $\mu$m region -. The stars are classified according to the shape of their SEDs and the properties of their IR excesses, as it is described in the two following sections. 

\subsubsection{Classification from the SED shape}
\label{Section:meeus}

Column 9 in Table \ref{Table:sample} shows the SED group according to the \citet{Meeus01} classification scheme. The Meeus group I and group II (M01 groups hereafter) SEDs have been associated with geometrical differences in the structure of the disks, flared and self-shadowed, respectively \citep{Meeus01,Dullemond02,Dullemond04}. Both groups show differences regarding the photopolarimetric behaviour \citep{Dullemond03}, presence and strength of the polycyclic aromatic hydrocarbons (PAHs) features \citep{AckeAncker04}, and grain growth as observed from sub-mm data \citep{Acke04}. The classification in M01 groups was made following the colour criterium in \citet{vanboekel03}.  This is based on the ratio L(nIR) (the integrated luminosity obtained from broad-band $JHKLM$ photometry) to L(IR) (the equivalent parameter, obtained from the IRAS 12, 25 and 60 micron fluxes), and on the non-colour corrected IRAS [12] -- [60] colour. This classification used the photometry given in Tables \ref{Table:photometry} and \ref{Table:photometrylong}, and agrees with previous ones reported for many stars in our sample \citep{Meeus01,Acke04,AckeAncker04,Acke10}. A direct visual inspection of Fig. \ref{Figure:SEDs} allows one to classify \object{HD 58647}, \object{BH Cep} and \object{NV Ori} as M01 group II sources, and \object{AS 442}, and \object{VY Mon} as group I. \object{T Ori} is provisionally classified as a group I source, but mid-IR photometry is necessary to confirm this result. Summarizing, 12 sources in our sample are classified as group I stars, the remaining 26 being group II objects.

\subsubsection{IR excess}
\label{Section:irexcess}

The sample was divided into two groups according to the shortest wavelength where the IR excess is apparent from the SED fits in Fig. \ref{Figure:SEDs}. Half of the stars show an IR excess starting at wavelengths corresponding to the $J$ or $H$ bands \citep[1.22 and 1.63 $\mu$m;][]{Bessel88}, and are included in the first group (group $JH$ hereafter). The objects in the second group (group $K$) show an IR excess starting at $\lambda$ $\geq$ 2.19 $\mu$m (filter $K$). The dust in the disks of group $JH$ sources is expected to extend up to the sublimation radius (around 0.5 AU, for a typical HAe star with T$_*$ = 9000 K and R$_*$ = 3 R$_{\odot}$)\footnote{The dust sublimation radius is R$_{sub}$ $\sim$ R$_*$(T$_{sub}$/T$_*$)$^{-2.1}$, with R$_*$, T$_*$ and T$_{sub}$ the stellar radius, effective temperature and sublimation temperature ($\sim$ 1600 K), respectively \citep{Robitaille06}.}. Assuming that the disk temperature radially declines following a R$^{-3/4}$ law \citep[see e.g.][and references therein]{Armitage09}, the small dust grains in group $K$ sources would be located at least twice far away from the central star. In particular, the SED of the four stars that show transition disks - defined as showing very weak or null excess up to $\sim$ 10 $\mu$m: \object{HD 34700}, \object{HD 141569}, \object{BO Cep}, and \object{51 Oph} - could be the signature of inner holes with sizes larger than the dust sublimation radius by an order of magnitude. Given that R$_{sub}$ increases with the temperature and radius of the central object, it is worth noting that the possible inner holes associated to HAeBe stars with IR excesses starting at the $K$ band could in principle have sizes similar to those associated to transitional T Tauri stars \citep[$\sim$ a few AU; see e.g.][]{Merin10}. Alternatively, the region in the disk sampled by an IR excess starting at a given wavelength is farther away from the central source in HAeBe stars than in TTs. The photometric band where the IR excess starts for each object is shown in Col. 10 of Table \ref{Table:sample}.

Almost all stars in the sample have measurements in $H$, $K$ and the IRAS bands, whose emission will be used to probe the disk at different radial distances. Column 11 in Table \ref{Table:sample} shows the intrinsic colour excess E($H-K$)$_0$ = ($H-K$)$_{dered}$ - ($H-K$)$_0$. This parameter measures the colour excess without the extinction contribution \citep{Meyer97}, reflecting the properties of the inner disk dust. The de-redenned colour, ($H-K$)$_{dered}$ = ($H-K$) + R$_V$E($B-V$) $\times$ (k$_K$/k$_V$ - k$_H$/k$_V$), k$_{\lambda}$ being the opacity at a given wavelength, is derived considering the extinction law in \citet{Robitaille07}, a total-to-selective extinction ratio R$_V$ = 5 \citep{Hernandez04}, and the $B$-$V$ colour excesses from Paper II. The intrinsic colours ($H-K$)$_0$ are from \citet{KenyonHartmann95}. Columns 12 and 13 of Table \ref{Table:sample} show the IRAS [12]-[25] and [25]-[60] colours, reflecting the warm mid-disk dust. The uncertainty in the IRAS colours - typically 0.15 magnitudes - is significantly larger than the redenning and intrinsic colour corrections \citep[see e.g.][]{PattenWillson91}, which were therefore not applied for these bands.

\subsection{Disk masses}
\label{Section:Mdisk}

Column 16 in Table \ref{Table:sample} shows the total (dust + gas) disk masses for the stars with mm photometry (see Table \ref{Table:photometryfar}). Estimates of the disk masses were derived following \citet{Beckwith90}, assuming that the emission is optically thin at mm wavelengths. The following expression, valid under the Rayleigh-Jeans regime, has been used:

\begin{equation}
\label{Eq:diskmasses}
M_{disk} \sim F_{\nu}\frac{\rm d^{2}}{\rm 2kT_{\rm D}}\frac{c^{2}}{\nu^{2}k_{\nu}},
\end{equation}
where F$_{\nu}$ is the flux measured at 1.3 mm for most stars in the sample, d is the distance, k is the Boltzmann constant, T$_{\rm D}$ is the dust temperature, and k$_{\nu}$ the opacity at the observed frequency. We adopted the standard k$_{\rm 1.3mm}$ = 0.02 cm$^{-2}$gr$^{-1}$ {(assuming a gas-to-dust ratio of 100)\footnote{The total (dust + gas) opacity is k$_{\rm 1.3mm}$ = k$_{\rm 1.3mm}^{dust}$/($r$+1), with k$_{\rm 1.3mm}^{dust}$ the dust opacity and $r$ the disk mass gas-to-dust ratio.} and a wavelength dependence k$_{\rm mm}$ $\propto$ $\lambda$ $^{-\beta}$ \citep{Beckwith90}, with 0 $<$ $\beta$ $\leq$ 2.5. The value of $\beta$ is representative of the maximum dust grain size, which is $\sim$ 2--2.5 for the IS medium and smaller for larger grains \citep{PollackHollenbach94,Draine06}. The parameters T$_{\rm D}$ and $\beta$ are those best fitting the available fluxes longward 350 $\mu$m from a graybody \citep[see e.g.][]{Andre93,Sandell00}. The input range for the dust temperatures was within $\sim$ $\pm$ 15 K around a central value taken from the relation between this parameter and the spectral type \citep{Natta00}. Figure \ref{Fig:SEDslong} shows the results from these fits. The typical $\beta$ value obtained from them is $\sim$ 1.0 (mean and median), pointing to grain growth with respect the interstellar medium. Several stars in the sample have one single mm measurement. For these objects we followed \citet{Natta00}, fixing $\beta$ to 1 and T$_{\rm D}$ to the corresponding value from the spectral type, according to their Table II. The final values for T$_{\rm D}$ and $\beta$ are listed in Cols. 14 and 15 of Table \ref{Table:sample}.

M$_{disk}$ ranges typically between 10$^{-4}$ and 0.4 M$_{\sun}$, 0.02 M$_{\sun}$ being the median value. Considering the stellar masses, the median value for the M$_{disk}$/M$_{*}$ ratio is 1 $\%$ \citep[i.e. the same as for lower-mass T Tauri stars; see][]{AndrewsWilliams07}. The extremely high disk mass for \object{LkHa 234} comes from the flux data set in \citet{PezzutoStrafella97} (see Table \ref{Table:photometryfar}), which is probably contaminated by environment emission. \citet{Fuente01} reported a flux of less than 20 mJy at 1.3 mm using a smaller 1.2'' $\times$ 1.3'' beam, deriving a disk mass  $<$ 0.1 M$_{\odot}$ for \object{LkHa 234}, whereas \citet{AlonsoAlbi09} reported M$_{disk}$ $<$ 0.2 M$_{\odot}$ from the same flux. It is finally noted that the uncertainties for M$_{disk}$ listed in Table \ref{Table:sample} only consider the errors in the mm fluxes. Therefore, they should be considered lower limits, since additional error sources are most probably coming from the T$_{\rm D}$ and $\beta$ values used in the fits, and specially from the uncertainty in the opacity (see  Section \ref{Section:discussion} and references therein). Gas-to-dust ratios significantly lower than the standard value of 100 \citep{Panic08,Thi10,Tilling12} could also affect the disk masses derived. For instance, our disk masses are an order of magnitude larger than those obtained from Eq. \ref{Eq:diskmasses} and a gas-to-dust ratio of 10.

\section{Correlations analysis}
\label{Section:analysis}

In this section we analyse possible correlations between the mass accretion rates from Paper II (Col. 5 in Table \ref{Table:sample}) and several properties of the stars in the sample, described in previous sections.

\subsection{Accretion rate and stellar age}
\label{Section:Macc_age}
The mass accretion rates are plotted against the stellar ages in Fig. \ref{Fig:age_Macc}. The accretion rate declines as the age increases, following a trend that can be fitted to an exponential expression $\dot{M}_{\rm acc}$($t$) = $\dot{M}_{\rm acc}$(0) $\times$ $\exp{(-t/\tau)}$, being $t$ the stellar age and $\tau$ the dissipation timescale \citep{Manoj06,Fedele10}. Apart from intrinsic scatter, the final value for $\tau$ is affected by uncertainties in the accretion rates and stellar ages, and also by upper and lower limits in both parameters for several stars. Both sources of error can be treated statistically in independent ways, i.e., there is no statistical technique able to deal simultaneosly with uncertainties for detections and upper/lower limit estimates, to our knowledge. We used the method in \citet{York04} to derive a fit assuming different ranges of reasonable uncertainties for both $\dot{M}_{\rm acc}$ and $t$, obtaining $\tau$ = 0.9$^{+1.0}_{-0.2}$ Myr. On the other hand, the astronomical survival analysis methods discussed in \citet{Isobe86} and \citet{Lavalley92} are useful to deal with censored data that include upper and lower limits. The {\tt schmidttbin} task in the {\tt iraf/stsdas} package is based on those works and provides a value of $\tau$ = 1.7 $\pm$ 0.6 Myr. Taking the average from both estimates, we assumed $\tau$ = 1.3$^{+1.0}_{-0.5}$ Myr as the accretion rate dissipation timescale for our sample.
 
\begin{figure}
\centering
\includegraphics[width=90mm,clip=true]{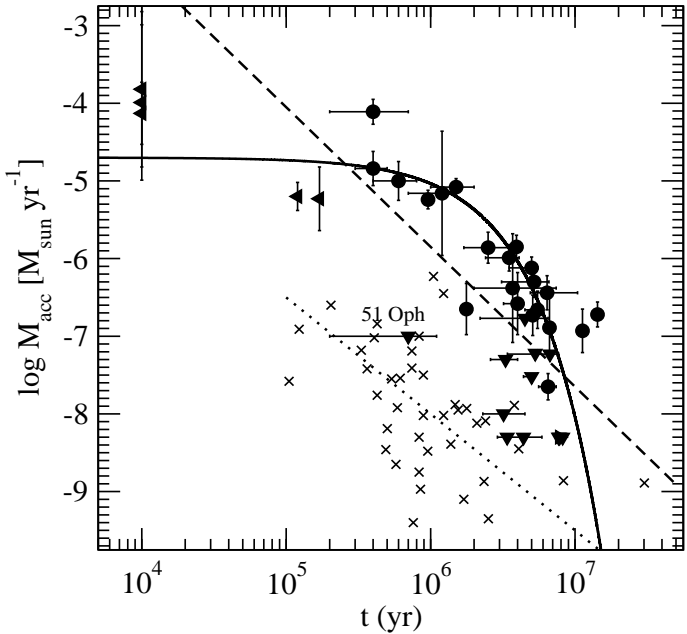}
\caption{Mass accretion rates versus stellar age for our HAeBes (solid symbols, upper and lower limits are represented with triangles) and T Tauri stars from \citet{Hartmann98} (crosses). The best fits for our sample are represented with solid and dashed lines, for the exponential ($\tau$ = 1.3$^{+1.0}_{-0.5}$ Myr) and power-law ($\eta$ = 1.8$^{+1.4}_{-0.7}$) expressions, respectively. \object{51 Oph} was not considered for the fit. The dotted line is the best power-law fit for T Tauri stars  \citep[$\eta$ $\sim$ 1.5; see][]{Hartmann98,Calvet00,Muzerolle00}.}
\label{Fig:age_Macc}
\end{figure} 

The accretion rate decrease with the stellar age can also be fitted by a power-law expression $\dot{M}_{\rm acc}$(t) $\propto$ t$^{-\eta}$, which is used in viscous dissipation models \citep{Hartmann98}. From the statistical methods explained above, our data can be reproduced with $\eta$ = 2.3$^{+0.9}_{-1.0}$, taking uncertainties for the ages and accretion rates into account. If the upper and lower limits are considered, $\eta$ = 1.2 $\pm$ 0.1. The average fit for our sample is $\eta$ = 1.8$^{+1.4}_{-0.7}$, which is shown in Fig \ref{Fig:age_Macc}, along with the best fit obtained for CTT stars with K and M spectral types from \citet{Hartmann98}.   

The exponential decay provides a better fit than the power-law one in terms of a lower $\chi$$^{2}$, and could reflect better the high but finite accretion rates of the youngest sources in the sample. However, this conclusion depends on the strong uncertainties for the accretion rates and ages of the youngest stars, which are also the most scarce in our sample. The stellar ages for \object{R Mon}, \object{VY Mon}, and \object{LkHa 234} are lower than 10$^4$ yr \citep{Manoj06}, and were derived from a different methodology and evolutionary tracks than for most star in the sample \citep{Montesinos09}. Those values may be not accurate, but they reflect the youthness of these objects, as is supported by other estimates, which provide ages $\leq$ 1 Myr for these stars \citep[see e.g.][]{AlonsoAlbi09,Tetzlaff11}. Our exponential fit provides a longer dissipation timescale than if their ages were older than assumed here. In this sense, the value for $\tau$ obtained in this work could be considered as an upper limit. Analogously, the power-law fit would be steeper if the objects mentioned are older. In addition, \object{51 Oph} stands apart from the trend, and was not considered in the fits. The evolutionary stage of this object is not clear, showing characteristics typical of both a HBe star and a MS star with a gas-rich debris disk \citep[][Paper II]{vandenAncker01,Stark09}. 

The exponential expression that fits the accretion rate decline with age for our sample provides a timescale shorter than that obtained for the H$\alpha$ equivalent width of HAeBe stars \citep[$\sim$ 3 Myr in][]{Manoj06}. However, the equivalent width itself does not measure the gas content and is subject to strong variability (see Paper I). The difference with the estimate by \citet{Manoj06} can be diminished by transforming the line equivalent widths into line luminosities, and then translating these into accretion rates (See Paper I and II). The CTT stars show a decrease of $\sim$ 1--2 dex in log $\dot{M}_{\rm acc}$ during the first Myr \citep{Fang09}, as our sample roughly show. Our value for $\tau$ is lower than the $\sim$ 2.3 Myr reported for K0--M5 pre-main sequence stars by \citet{Fedele10}, although they are comparable considering the large uncertainties. Those authors made a statistical study of the objects that show signs of accretion in star forming regions at different ages, concluding that planet formation and/or migration in the inner disk might be a viable mechanism to halt further accretion onto the central star on such a short timescale. Indeed, core-accretion models \citep{Pollack96,Mordasini08} require $\sim$ 3 Myr including planet migration \citep{Alibert04,Alibert05}, while gravitational instability \citep{Durisen07,Boss11} needs 10$^3$-10$^6$ yr to form planets. The fast inner gas dissipation timescale suggested from our sample is more consistent with gravitational instability models, and contrasts with that required for the formation of planets from core-accretion ones. Other processes apart from planet formation might contribute to explain the fast decrease of the accretion rate in our stars, as is discussed in following sections.

The analysis of the accretion rate decline represented as a power-law also points to a slightly faster inner disk gas dissipation in HAeBes, given that the mean $\eta$ value derived for our sample is higher than that for CTT stars. Despite the strong uncertainties, $\eta$ $\sim$ 1.5 seems more consistent for low-mass objects \citep{Hartmann98,Calvet00,Muzerolle00}, and the most recent estimates indicate an even slower mass accretion rate decay \citep[$\eta$ $\sim$ 1.2;][]{Aurora10,Caratti12}.

We have followed a similar procedure as used for CTT stars by directly comparing the $\dot{M}_{\rm acc}$ and $t$ values in the HAeBe regime. In this way, we have shown that the accretion rate and the stellar age are strongly correlated in our sample (Fig. \ref{Fig:age_Macc}), with a Spearman probability of false correlation of only 8.5$\times$10$^{-6}$ $\%$. However, the interpretation above of this trend also has to consider a caveat that comes from internal dependences on the stellar mass. First, $\dot{M}_{\rm acc}$ depends on it as $\sim$ M$_*^5$ for our sample (Paper II). Second, the age of HAeBe stars is also correlated with M$_*$, the youngest objects being the most massive \citep[][Paper II]{vanboekel05}. The partial correlations technique \citep[e.g.][]{Wall03} provides the probability of false correlation between two variables, with the effect of a third controlling variable removed. The Spearman probability of false correlation between $\dot{M}_{\rm acc}$ and $t$ for our sample in Fig. \ref{Fig:age_Macc} rises to 85 $\%$, if the dependences of both $\dot{M}_{\rm acc}$ and $t$ on M$_*$ are considered. As a consequence, there is a high probability that the accretion rate decline in Fig. \ref{Fig:age_Macc} is driven by the dependence on the stellar mass. Therefore, the $\tau$ and $\eta$ values describing this evolution should be viewed with caution. These results suggest that the evolution of the accretion rate should be studied by dividing the HAeBe regime into small stellar mass bins, which would therefore need larger samples than studied here. This division would also be physically consistent, given that different evidence suggests that the accretion paradigm could change at some point within the HAeBe regime, the less massive objects showing properties more similar to CTT stars \citep[][Paper I, Paper II]{Vink02,Eisner04,Mottram07}. Almost all stars in our sample with M$_*$ $>$ 2.5 M$_{\odot}$ are younger than 1 Myr, which could indicate that at older ages most of these objects have already reached the MS phase. The faster evolution of the massive HAeBes would agree with recent observational works; in particular, \citet{Roccatagliatta11} reported that sources with masses $>$ 2 M$_{\odot}$ in the OB association IC 1795 have a disk fraction of 20$\%$, while lower mass objects (2--0.8 M$_{\odot}$) have a disk fraction of 50$\%$.  

\subsection{Accretion rate and spectral energy distribution}
\label{Section:Macc_sed}

Figure \ref{Fig:hist_evol_meeus} shows the distribution of mass accretion rates for the whole sample, as well as for the M01 group I and group II sources.  The typical - median - accretion rate for each group is $\sim$ 10$^{-7}$ M$_{\sun}$ yr$^{-1}$, the same as for the whole sample. Group II stars show accretion rates distributed roughly homogeneously around the median. Group I sources show more scatter, but their accretion rates also cover the full possible range for the HAeBe regime. Three group I sources, namely \object{R Mon}, \object{VY Mon} and \object{LkHa 234}, are hot -T$_{*}$ $>$ 12000 K- HAeBes with strong Balmer excesses most probably associated with high accretion rates (Paper II). Even considering these objects, the null hypothesis that both M01 groups are drawn from the same accretion rate distribution is supported by a two-sample Kolmogorov-Smirnov (K-S) test with a probability of 75$\%$.

\begin{figure}
\centering
\includegraphics[width=90mm,clip=true]{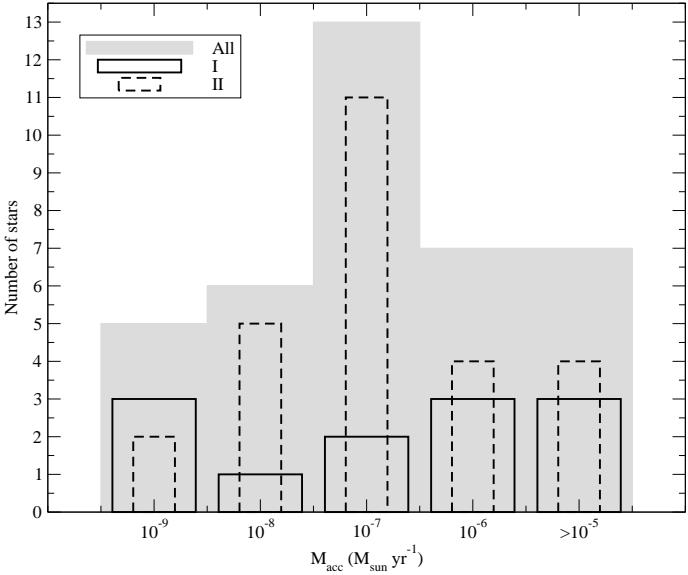}
\caption{Histogram showing the mass accretion rates for the 38 stars in the sample (shaded bars), and the stars belonging to the groups I and II (12 and 26 objects; solid and dashed lines) from the classification of \citet{Meeus01}. The probability that both groups are drawn from the same accretion rate distribution is 0.75.}
\label{Fig:hist_evol_meeus}
\end{figure}

A larger number of M01 group II stars, compared to group I, is found in all works dealing with this classification (see Sect. \ref{Section:meeus} and references below). The origin of this division could be understood by arguing that objects that show flared group I SEDs are in a younger, faster evolving stage than self-shadowed group II stars. This view, which would point to an evolution from group I to group II sources, is theoretically and observationally supported \citep{Dullemond02,Dullemond04,Acke04}. It would imply that group I sources should show higher accretion rates than group II stars. As we mentioned, such a conclusion cannot be drawn from our sample, and the implications deserve more study. On the other hand, our data agree with previous works that reported larger dust grain sizes in group II stars \citep{Acke04} - our median $\beta$ value obtained from the fits in Fig. \ref{Fig:SEDslong} is 0.6 for this group, against 1.4 for group I sources -, and with works that reported that UXOrs tend to be group II sources \citep{Dullemond03} - \object{LkHa 234} is the only UXOr \citep{Oudmaijer01,Rodgers03} in our sample not belonging to that group, but this should be taken with caution given that its IR photometry is likely contaminated by the environment \citep{Eiroa98} -. In addition to the interpretation in \citet{Dullemond03}, the fact that group II sources include most UXOrs could be related to an inclination effect \citep{Natta97,NattaWhitney00}, which in turn could be affecting the SED shape classification \citep{Robitaille07,Meijer08}. 

From the observational perspective, the absence of a clear correlation between the accretion rates and the M01 groups was somehow expected. The M01 classification scheme strongly depends on the shape of the SED at long wavelengths \citep[in particular on the IRAS 12, 25 and 60 $\mu$m fluxes; see][and Sect. \ref{Section:meeus}]{vanboekel03}, and Fig. \ref{Fig:Macc_IRAS} shows that the mass accretion rate is not correlated with the colours at these bands. 

\begin{figure}
\centering
\includegraphics[width=90mm,clip=true]{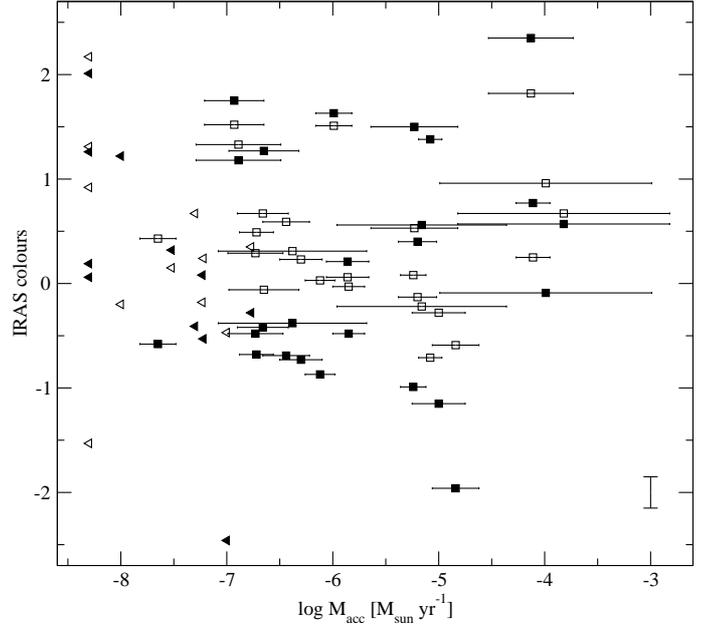}
\caption{IRAS colours [12]-[25] (open symbols) and [25]-[60] (filled symbols) versus the mass accretion rate. The typical uncertainty for the colours is plotted close to the bottom right corner. Triangles are upper limits for the accretion rates.} 
\label{Fig:Macc_IRAS}
\end{figure}

The contribution from the accretion shock is stronger at shorter wavelengths (see Paper II and references therein), and could be measurable up to nIR wavelengths \citep{Muzerolle04}. Figure \ref{Fig:Macc_nIR} shows the intrinsic colour excess E($H-K$)$_0$ against the mass accretion rates, along with the corresponding values for TT stars. Both parameters are correlated, with a Spearman probability of false correlation of 1.5 $\times$ 10$^{-5}$ $\%$ for our sample. However, this relation becomes not statistically significant for $\dot{M}_{\rm acc}$ $\lesssim$ 1.5 $\times$ 10$^{-7}$ M$_{\sun}$ yr$^{-1}$. The slope for stronger accretion rates is 0.12 $\pm$ 0.05, equal within the uncertainties to that for CTT stars \citep[0.10 $\pm$ 0.05;][]{Meyer97}. Both samples can be fitted simultaneously with a linear expression with slope 0.15 $\pm$ 0.03. 

\begin{figure}
\centering
\includegraphics[width=90mm,clip=true]{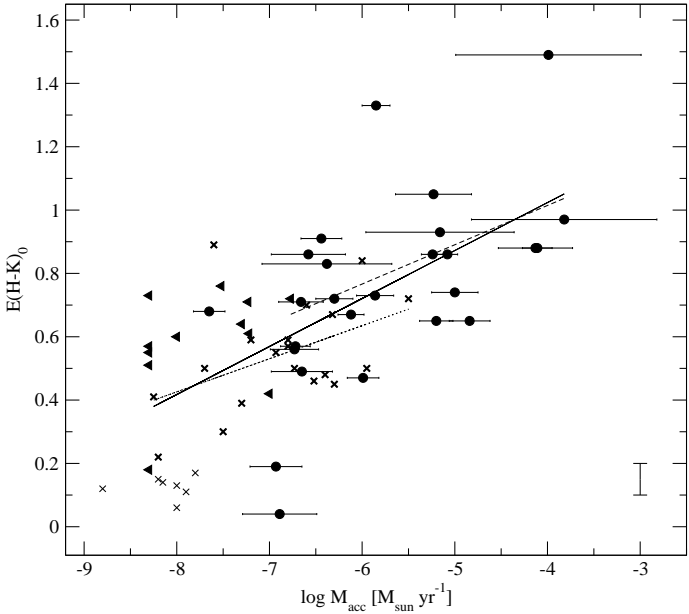}
\caption{Intrinsic colour excess in $H-K$ versus the mass accretion rate. The typical uncertainty for the colour is plotted close to the bottom right corner. Triangles are upper limits for the accretion rates. Crosses represent classical T-Tauri (boldface crosses) and weak T Tauri (normal crosses) stars from \citet{Meyer97}. The best fits are E($H-K$)$_0$ = 1.51($\pm$0.26) + 0.12($\pm$0.05) $\times$ log $\dot{M}_{\rm acc}$ (dashed line, Pearson correlation coefficient 0.50) for HAeBes with $\dot{M}_{\rm acc}$ $>$ 1.5 $\times$ 10$^{-7}$ M$_{\sun}$ yr$^{-1}$, E($H-K$)$_0$ = 1.26($\pm$0.34) + 0.10($\pm$0.05) $\times$ log $\dot{M}_{\rm acc}$ (dotted line, Pearson correlation coefficient 0.50) for CTTs, and E($H-K$)$_0$ = 1.63($\pm$0.17) + 0.15($\pm$0.03) $\times$ log $\dot{M}_{\rm acc}$ (solid line, Pearson correlation coefficient 0.65) for both HAeBes showing high accretion rates and CTTs.} 
\label{Fig:Macc_nIR}
\end{figure}

The similar correlation between the mass accretion rates and the $H$-$K$ colour excesses followed by T Tauri and HAeBe stars points to a common origin for both trends, suggesting that the inner dust of HAeBe stars is not only heated by re-processing light from the central star, but also by the contribution of accretion. Because the relation between the mass accretion rate and the nIR colour breaks for $\dot{M}_{\rm acc}$ $\lesssim$ 1.5 $\times$ 10$^{-7}$ M$_{\sun}$ yr$^{-1}$, this value can be considered as an approximate limit to divide passive and active disks in HAeBe stars \citep[see also][]{vandenancker05}. This is one order of magnitude larger than the corresponding accretion rate that divides weak and classical T Tauri stars (Fig. \ref{Fig:Macc_nIR}), as can be estimated by assuming viscous accretion dissipation and stellar irradiation as the only heating sources of the disk \citep{Calvet00}:
\begin{equation}
\label{Eq:Mdiv}
\dot{M}_{\rm pa} \sim 2 \times 10^{-8} M_{\sun}  yr^{-1} \left(\frac{T_{*}}{4000 K}\right)^{4} \left(\frac{R_{*}}{2R_{\sun}}\right)^{3} \left(\frac{M_{*}}{0.5M_{\sun}}\right)^{-1},
\end{equation}
where $\dot{M}_{\rm pa}$, the accretion rate dividing passive and active disks, is derived assuming that the mass accretion rate through the inner disk is constant and equal to the rate at which mass is transferred onto the central object. Applying Eq. \ref{Eq:Mdiv} to a typical HAe star (T$_{*}$ = 9000 K, M$_{*}$ =  3 M$_{\sun}$, R$_{*}$ = 3 R$_{\sun}$), the observational value 1.5 $\times$ 10$^{-7}$ M$_{\sun}$ yr$^{-1}$ is also recovered. This corresponds to log (L$_{acc}$/L$_{\odot}$) $\sim$ 0.60, which, along with the empirical relations with the emission line luminosities provided in Paper II, indicates that active disks around HAeBes tend to show observational thresholds log (L$_{H\alpha}$/L$_{\odot}$) $>$ -1.5, log (L$_{[\ion{O}{i}]6300}$/L$_{\odot}$) $>$ -3.7, and log (L$_{Br\gamma}$/L$_{\odot}$) $>$ -3.2.   

Figure \ref{Fig:hist_evol_JHK} shows the distribution of mass accretion rates according to the classification from the wavelength where the IR excess starts (groups $JH$ and $K$). In this case, the K-S test provides a probability of only 1.8$\%$ that both groups follow the same accretion rate distribution (null-hypothesis), against the alternative that both groups show different accretion rate distributions. Moreover, if the alternative hypothesis is that the stars in the group $JH$ have a smaller cumulative distribution function than the remaining objects, then the null hypothesis is rejected with a probability of 99.1 $\%$. Therefore, we conclude that there is a significant trend for the stars with signs of inner dust dissipation to have lower accretion rates ($\leq$ 10$^{-7}$ M$_{\sun}$ yr$^{-1}$), being the objects with IR excesses starting at $J$ or $H$ the strongest accretors. The typical - median - mass accretion rate for these stars is $\sim$ 10$^{-6}$ M$_{\sun}$ yr$^{-1}$, one order of magnitude higher than that for the group $K$ objects. This difference reaches two orders of magnitude if mean values are adopted instead of medians. Some stars deviate from the general trend that relates the accretion rates and the wavelength where the IR excess starts. For instance, \object{HD 141569} shows an SED typical of a transitional disk (see the corresponding panel in Fig. \ref{Figure:SEDs}), pointing to significant inner dust dissipation and/or grain growth. At the same time, there are no signs of significant inner gas dissipation, since its mass accretion rate is only slightly lower than the median value for the sample (1.3 $\times$ 10$^{-7}$ against 2.4 $\times$ 10$^{-7}$ M$_{\sun}$ yr$^{-1}$).

\begin{figure}
\centering
\includegraphics[width=90mm,clip=true]{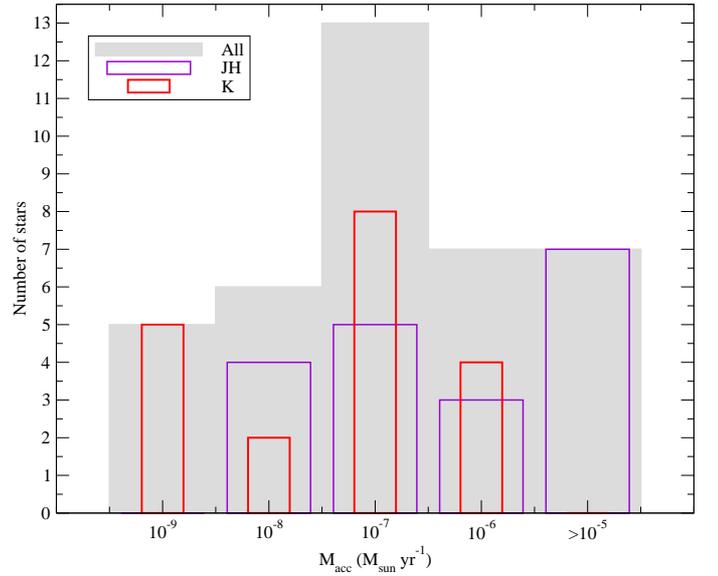}
\caption{Histogram showing the mass accretion rates for the 38 stars in the sample (shaded bars), and the stars belonging to the groups $JH$ and $K$ (half of the stars in each group; violet and red lines). The probability that both groups are drawn from the same accretion rate distribution is $<$ 0.018.}
\label{Fig:hist_evol_JHK}
\end{figure}

Finally, the median $\beta$ value obtained from the fits of the mm fluxes (Fig. \ref{Fig:SEDslong}) has a similar value, about 1.1, for both groups $JH$ and $K$, which contrasts with the results mentioned above about the differences for the M01 groups. This suggests that a lower nIR emission is is not necessarily linked to grain growth as determined by mm measurements; these sample the outer disk, and are more closely related to the possible decrease of the disk flaring angle associated to M01 group II sources.

\subsection{Accretion rate and disk mass}
\label{Section:Macc_Mdisk}

Figure \ref{Fig:Mdisk_Macc} shows the mass accretion rate against the disk mass for the stars in our sample, as well as for T Tauris. The Spearman probability of false correlation between log $\dot{M}_{\rm acc}$ and log M$_{disk}$ is 2.9 $\%$ for our sample, which is best fitted from log $\dot{M}_{\rm acc}$ = 1.1($\pm$0.3) $\times$ log M$_{disk}$ - 5.0($\pm$0.5), with a Pearson correlation coefficient of 0.60. Most stars in the sample deviate from that expression by less than one order of magnitude (dashed lines in Fig. \ref{Fig:Mdisk_Macc}). The slope for T Tauris is also close to unity, once they are divided into transition and non-transition disks \citep{Najita07}, which is the value expected from simple viscous dissipation models \citep[see][]{Hartmann98,Dullemond06,Najita07}. The trend could break for HBe stars more massive than those studied here, given the uncertainty of their accretion rates, and because some of their disk masses are lower than those for T Tauri stars \citep{AlonsoAlbi09}. 

\begin{figure}
\centering
\includegraphics[width=90mm,clip=true]{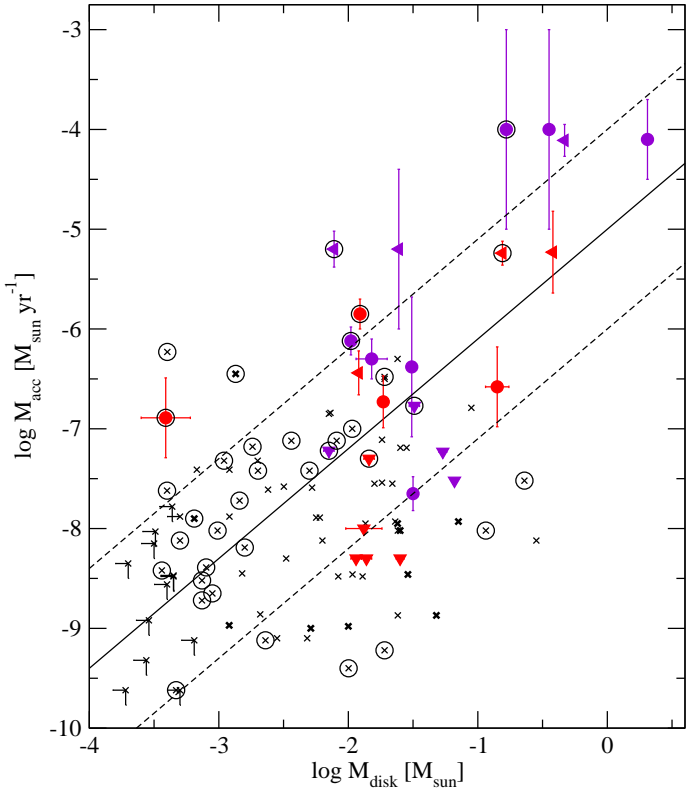}
\caption{Mass accretion rate against disk mass. Crosses are T Tauri stars from \citet{Hartmann98} and \citet{Najita07}. Dark crosses are transitional disks. The values for our 26 HAeBes with disk mass estimates are represented with solid symbols, triangles for the upper limits. Group $JH$ stars are plotted in violet and group $K$ stars in red. The solid line represents the best fit to our sample and the dashed lines $\pm$ 1 dex. The close binaries for both T Tauri and HAeBe stars are surrounded by solid circles.}
\label{Fig:Mdisk_Macc}
\end{figure}

Given that M$_{disk}$($r$')/M$_{disk}$($r$) = ($r$'+1)/($r$+1), with M$_{disk}$($r$) the total disk mass from Eq. \ref{Eq:diskmasses} for a gas-to-dust ratio $r$, a different value $r$' would shift the data in the x-axis of Fig. \ref{Fig:Mdisk_Macc}. For instance, if $r$' = 10, instead of the assumed value $r$ = 100, then the x values would shift 1 dex to the left but the $\dot{M}_{\rm acc}$--M$_{disk}$ relation would still have a slope close to 1. In conclusion, the slope of this trend should not be affected by the chosen value of the gas-to-dust ratio, if this is roughly the same for all objects.  

Our sample is divided in Fig. \ref{Fig:Mdisk_Macc} according to the wavelength where the IR excess starts, and considering the  close binaries separated by less than 1000 AU, in analogy with the study in \citet{Najita07} for lower-mass stars in Taurus. In this case it was reported that stars with transition disks, when compared with non-transitional ones, tend to show accretion rates ten times lower for a given disk mass, and median disk masses about four times larger. In addition, close binary stars were reported to show slightly lower median disk masses than non-binaries. Although most of our group $K$ stars are too far from being transitionals, from the definition considering the wavelength where the IR excess is apparent, two of the three mentioned conclusions from \citet{Najita07} also apply to the HAeBes. First, the mass accretion rates tend to be lower by a factor $\gtrsim$ 10 for our stars with IR excesses starting at longer wavelengths (Sect. \ref{Section:Macc_sed}). Second, the close binaries in our sample have lower median disk masses compared to systems without such companions (1.2 $\times$ 10$^{-2}$ against 3.1 $\times$ 10$^{-2}$ M$_{\sun}$). However, in contrast with the results in \citet{Najita07}, the disk masses of the HAeBes with signs of inner disk dissipation are typically three times lower (median values) than those with IR excesses measured at the $J$ or $H$ bands - considering the uncertainties in Table \ref{Table:sample}, the mean disk mass for group $JH$ is 0.24 $\pm$ 0.02 M$_{\odot}$, while that for group $K$ is only 0.066 $\pm$ 0.003 M$_{\odot}$. This difference between TTs and HAeBes could be pointing to different physical mechanisms dominating inner disk clearing in both regimes (see the following section). 

\section{Discussion}
\label{Section:discussion}

The similarities found when the accretion rate is related to the nIR colours (Sect. \ref{Section:Macc_sed}), and disk masses (Sect. \ref{Section:Macc_Mdisk}), point to a common origin that explains these trends for both the T Tauri and the intermediate-mass regime. The relation with the nIR colour excess extends the corresponding one for TTs to HAeBe stars (Fig. \ref{Fig:Macc_nIR}). Both trends can be understood if the dust of the inner disk is heated not only by reprocessed stellar light, but also by the viscous accretion contribution \citep{Calvet00}. The slope relating the accretion rate and the disk mass is close to unity (Fig. \ref{Fig:Mdisk_Macc}), which is also expected from simple viscous disk models \citep[e.g.][]{Dullemond06}. 

Apart from the analogies mentioned, we found two major differences when our data are related to the corresponding for T Tauri stars. First, our results are consistent with a slightly shorter inner gas dissipation timescale (Sect. \ref{Section:Macc_age}). As argued in that section, this could have implications on the physical mechanism able to form planets around stars more massive than the sun on such a shorter timescale. Second, the relative position of HAeBe stars with signs of inner dust dissipation in the $\dot{M}_{\rm acc}$--M$_{disk}$ plane (Fig. \ref{Fig:Mdisk_Macc}), compared to the remaining HAeBes without signs of disk clearing, differs from a similar comparison between transition and classical disks around TTs. In particular, our group $K$ sources show lower accretion rates and disk masses than group $JH$ sources, whereas transitional T Tauri stars show lower accretion rates but larger disk masses than classical TTs \citep{Najita07}. The $\dot{M}_{\rm acc}$--M$_{disk}$ plane is a potential tool for studying the physical mechanisms that dominate circumstellar disk dissipation. In summary \citep[see][for more details]{Alexander07,Najita07}, large disk masses and low mass accretion rates, compared to those typically shown by non-transitional disks, are expected if Jovian planets are formed around stars with signs of inner dust dissipation. Large disk masses are necessary to form this type of planets, and at the same time, the inner gas content - and therefore the accretion rate - is strongly diminished since part of the gas otherwise destined to the central star accretes onto the planets. In contrast, photoevaporated disks are in principle more consistent with both low disk masses and accretion rates. It is unknown if this type of conclusions from the modelling in \citet{Alexander07} could be applied to stellar masses higher than 1M$_{\odot}$. As we mentioned (see also Sect. \ref{Section:Macc_Mdisk}), most of our stars with signs of inner dust dissipation have mass accretion rates and disk masses lower than the corresponding median values for the remaining sources by a typical factor of $\sim$ 10 and 3, respectively. Some examples are \object{HD 34282}, \object{HD 34700}, \object{HD 142666} or \object{BF Ori}, as well as transition disks associated with distant and close binary stars (\object{HD 34700} and \object{HD 141569}, respectively). The lower values for both $\dot{M}_{\rm acc}$ and M$_{disk}$ could be expected from photoevaporation sweeping out both the circumstellar gas and dust content \citep{Clarke01,AlonsoAlbi09}. Indeed, from the theoretical point of view photoevaporation could be more relevant for dissipating disks in the HAeBe regime than in the T Tauri one \citep[see e.g.][]{Takeuchi05}. On the other hand, \object{T Ori} is the only star in our sample with signs of inner dust dissipation that shows a lower mass accretion rate and a higher disk mass than the corresponding median values for the stars with an IR excess starting at $J$ or $H$ bands. As argued by \citet{Najita07} for the transition disks in Taurus, those properties are anticipated by several planet formation theories, suggesting that Jovian planets could be playing a significant role in explaining the inner disk dissipation of \object{T Ori}. The possible presence of a T-Tauri type stellar companion located at $\sim$ 3500 AU, considering a distance to the system of 470 pc \citep{Leinert97,Montesinos09}, could also be affecting the position of \object{T Ori} in the $\dot{M}_{\rm acc}$--M$_{disk}$ plane. \object{CQ Tau} shows similar characteristics to \object{T Ori}, with signs of inner disk dissipation, a low value of $\dot{M}_{\rm acc}$ and a high M$_{disk}$, compared with the medians for the whole sample. Thus, the formation of Jovian planets could also be playing a role for this star. Grain growth is also apparent from the fitted value of $\beta$ ($\sim$ 0.6). 

In conclusion, we have provided several observational indications that could be consistent with various disk dissipation mechanisms - viscous accretion, photoevaporation, planet formation, grain growth -. The dominating mechanism could differ depending on each object. To establish whether there could be a dominating disk dissipation mechanism for the whole HAeBe regime would require additional observational and theoretical effort.

The discussion above must take into account that disk masses estimated from mm measurements could be affected by strong uncertainties. For a sample of T Tauri stars, \citet{AndrewsWilliams07} found that disk masses from mm continuum emission underestimate those from accretion by roughly one order of magnitude. Following \citet{Hartmann98}, the amount of material accreted from t$_0$ to t$_{MS}$ - with t$_0$ the age of the star and t$_{MS}$ the time to reach the main sequence - is a lower limit for the disk mass at t$_0$, and is given by

\begin{equation}
\label{Eq:Mdisk}
M_{disk} \geq \int_{t_0}^{t_{MS}} \dot{M}_{\rm acc}(t) dt.
\end{equation}
Assuming $\dot{M}_{\rm acc}$(t) = At$^{-\eta}$ + B, with $\eta$ $>$ 1, A, B constants that can be determined from a measurement of the mass accretion rate at a given t$_0$, and under the condition $\dot{M}_{\rm acc}$(t$_{MS}$) $\sim$ 0, we derive
\begin{equation}
\label{Eq:MdiskMacc}
M_{disk} \geq \left(\frac{\dot{M}_{\rm acc}(t_0)}{t_0^{-\eta} - t_{MS}^{-\eta}}\right) \times \left(\frac{1}{\eta - 1}\right) \times \left(\frac{1}{t_0^{\eta - 1}} - \frac{1}{t_{MS}^{\eta - 1}} + \frac{t_0(\eta - 1)}{t_{MS}^{\eta}}\right),
\end{equation}
which for t$_{MS}$ $\gg$ t$_0$, yields   
\begin{equation}
\label{Eq:MdiskMacc2}
M_{disk} \geq \frac{\dot{M}_{\rm acc}(t_0)}{\eta - 1} \times t_0.
\end{equation}

\begin{figure}
\centering
\includegraphics[width=90mm,clip=true]{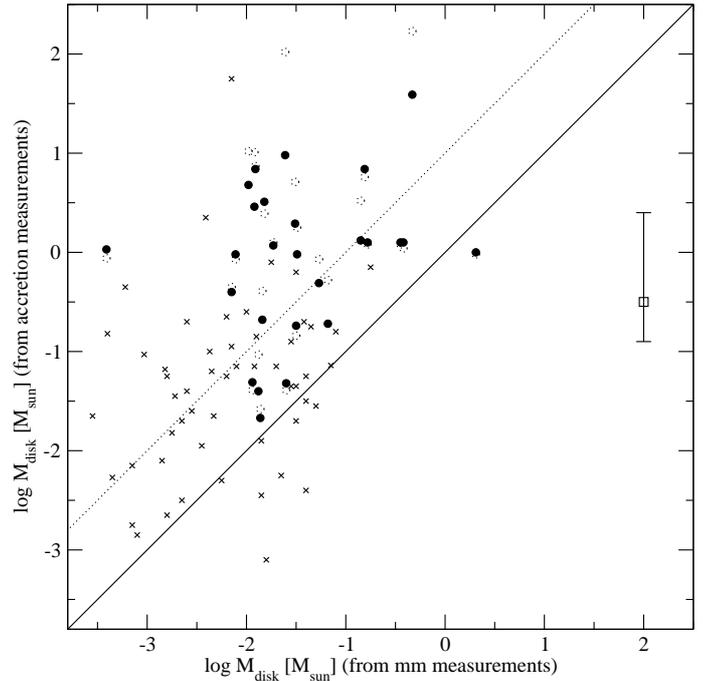}
\caption{Disk masses from accretion versus those from dust emission at mm wavelengths. Crosses are T Tauri stars taken from \citet{AndrewsWilliams07} and circles the estimates for our sample. The solid line represents equal values and the dotted line bisects both samples, indicating that disk masses from accretion are typically one order of magnitude larger than those from mm measurements. The filled circles are derived from Eq. \ref{Eq:MdiskMacc2} and $\eta$ = 1.8. The error bar at the right side represents the uncertainty coming from using $\eta$ = 1.1 (upper limit) or 3.2 (lower limit). The open-dotted circles are the estimates from Eq. \ref{Eq:MdiskMacc}, $\eta$ = 1.8, and the theoretical times to reach the main sequence from \citet{Tayler94}.}     
\label{Fig:diskcomp}
\end{figure}

Equation \ref{Eq:MdiskMacc2} was applied by \citet{Hartmann98} and \citet{AndrewsWilliams07} to T-Tauri stars. It gives practically equal values to Eq. \ref{Eq:MdiskMacc} for these objects, and for many of our stars, but Eq. \ref{Eq:MdiskMacc} provides disk masses that can be one order of magnitude larger than those from Eq. \ref{Eq:MdiskMacc2} for the most massive stars in our sample. These objects are expected to evolve very fast because their ages are similar to the theoretical times to reach the main sequence \citep[i.e. t$_0$ $\sim$ t$_{MS}$, see Fig. 4 in paper II and e.g.][]{Tayler94}. Figure \ref{Fig:diskcomp} compares the disk masses derived from the mass accretion rate to those from the cold dust emission. The results for our sample extend the findings for T Tauri stars, indicating that disk masses inferred from mm measurements are smaller than those from accretion by an order of magnitude on average. Following \citet{AndrewsWilliams07}, this may be the result of particle growth (up to approximately meter sizes), which leads to an overestimate of the opacity used to compute M$_{disk}$ from the millimeter luminosity. This view is supported by the low $\beta$ values derived for most stars in our sample with SED fitting at mm and sub-mm wavelengths (Fig. \ref{Fig:SEDslong}), which are similar to those for T Tauris \citep{Ricci10} and point to substantial grain growth in the CS environment of pre-main sequence stars \citep[see][and references therein]{AndrewsWilliams07}. For a fixed value of the dust opacity, the use of Eq. \ref{Eq:diskmasses} with a gas-to-dust ratio $\sim$ 1000 yields disk mass estimates an order of magnitude higher, which would roughly match those derived from accretion. Although there is no strong evidence for such a high ratio compared to what is typical for the interstellar medium, this possibility cannot be excluded in some objects \citep[see e.g.][]{Martin07,Carmona08}.

\section{Summary and conclusions}
\label{Section:conclusion}

We have related the mass accretion rates with the stellar ages, SED properties, and disk masses of a sample of 38 F-G, HAe and late-type HBe PMS objects. We have compared the trends with the corresponding for lower mass T Tauri stars. Our main results and conclusions are:

\begin{itemize}
 \item The mass accretion rate decreases with the stellar age. This decline can be fitted to  $\dot{M}_{\rm acc}$(t) $\propto$ $\exp{(-t/\tau)}$ or to $\dot{M}_{\rm acc}$(t) $\propto$ t$^{-\eta}$. The first expression provides a dissipation timescale $\tau$ = 1.3$^{+1.0}_{-0.5}$ Myr. The second expression is the same as used in simple viscous dissipation models for CTT stars, and yields $\eta$ = 1.8$^{+1.4}_{-0.7}$. The large error bars reflect intrinsic scatter, uncertainties and upper-lower limit estimates for the accretion rates and stellar ages. The values obtained should be viewed with caution, given that both the the stellar age and the accretion rate also depend on the stellar mass. However, our expressions point to a slightly faster dissipation of the inner gas in HAeBes, when compared to lower mass T Tauri stars.  
  \item Our sample does not show any apparent correlation between $\dot{M}_{\rm acc}$ and the SED shape according to the \citet{Meeus01} classification scheme. This is more closely related to the grain size or the UXOr behaviour, as was previously reported. 
 \item The correlation between the near-IR excess and the accretion rate extends from the CTT to the intermediate-mass regime. The mass accretion rate that divides active and passive disks rises one order of magnitude for our sample ($\sim$ 2 $\times$ 10$^{-7}$ M$_{\sun}$ yr$^{-1}$), as can be inferred assuming viscous accretion and stellar irradiation as the only heating sources of the inner disk. 
 \item In agreement with simple viscous disk models, the mass accretion rate correlates with the disk mass as $\dot{M}_{\rm acc}$ $\propto$ M$_{disk}$$^{1.1(\pm0.3)}$, with a scatter lower than $\pm$ 1 dex for most of the stars, which is roughly followed by CTT stars once they are classified into transition and non-transition disks \citep{Najita07}.
 \item Most stars in our sample with signs of inner dust disk clearing (IR excesses starting at the $K$ band or longer wavelengths) show lower accretion rates and disk masses than those for the remaining objects. This contrasts with previous results for transitional T Tauri stars, and could suggest that a different physical process - such as photoevaporation - plays a major role in dissipating disks in HAeBes. 
\end{itemize}

Finally, we extended the previously reported mismatch between the disk masses derived from cold dust emission and accretion to the HAeBe regime. More efforts are necessary to break that inconsistency, which is most probably related to the way that disk masses are estimated from mm measurements. 

 \begin{acknowledgements}
The authors thank the anonymous referee for his/her useful comments on the original manuscript, which helped us to improve the paper. C. Eiroa, G. Meeus, I. Mendigut\'{\i}a, and B. Montesinos are partially supported by grant AYA-2008 01727. This research made use of the SIMBAD database, operated at the CDS, Strasbourg, France 
\end{acknowledgements}

\clearpage

\appendix
\section{Tables with photometry and SEDs}
\label{appendix}

\begin{table}[!htbp]
\caption{Ultraviolet, optical and near-IR photometry}
\label{Table:photometry}
\centering
\begin{tabular}{lrrrrrrrrrrr}
\hline\hline
Star&IUE&U&B&V&R&I&J&H&K&L&M\\ 
\hline 
HD 31648 &$\surd$&7.84&7.78&7.62&7.46&7.43&6.98&6.41&5.67&4.61$^{1}$&...\\  
HD 34282 &...&10.19&10.03&9.87&9.77&9.63&9.13&8.55&7.93&6.25$^{1}$&6.16$^{1}$\\
HD 34700 &$\surd$&9.75&9.67&9.09&8.70&8.51&7.96&7.63&7.40&6.89$^{2}$&...\\
HD 58647 &$\surd$&\emph{6.81}$^{*}$&6.92&6.85&6.73&6.73&\emph{6.44}$^{*}$&\emph{6.11}$^{*}$&\emph{5.44}$^{*}$&...&...\\
HD 141569&$\surd$&7.22&7.20&7.09&7.00&7.04&6.66&6.58&6.51&6.69$^{3}$ &6.50$^{3}$\\ 
HD 142666&$\surd$&9.41&9.16&8.68&8.31&8.01&7.30&6.74&6.09&5.05$^{3}$ &4.85$^{3}$\\
HD 144432&$\surd$&8.47&8.46&8.14&7.82&7.53& \emph{7.07}& \emph{6.53}&\emph{5.93}&5.00$^{3}$ &4.52$^{3}$\\
HD150193 &$\surd$&9.71&9.39&8.89&8.41&7.81&\emph{6.89}&\emph{6.14}&\emph{5.33}&4.19$^{4}$&3.76$^{4}$\\
HD163296 &$\surd$&6.95&6.92&6.86&6.73&6.67&6.17&5.49&4.71&3.51$^{4}$&3.13$^{4}$\\
HD 179218&$\surd$&7.55&7.46&7.38&7.25&7.21&6.99&6.61&5.95&4.68$^{1}$&4.18$^{1}$\\
HD 190073&...&\emph{8.46}$^{1}$&7.86&7.73&7.66&7.65&7.22&6.63&5.85&4.58$^{1}$&...\\
AS 442&...& \emph{11.90}$^{5}$&11.56$^{*}$&10.90$^{*}$&\emph{10.18}$^{5}$&\emph{9.95}$^{6}$&8.65$^{*}$&7.70$^{*}$&6.62$^{*}$&...&...\\
VX Cas   &$\surd$&11.83&11.65&11.39&11.19&10.94&10.08&9.12&8.15&...&...\\
BH Cep&...& 12.18&11.91&11.23&10.81&10.34&\emph{9.69}$^{*}$&\emph{8.99}$^{*}$&\emph{8.31}$^{*}$&...&...\\
BO Cep&...&11.86&11.97&11.52&11.12&10.71&\emph{10.32}$^{*}$&\emph{9.85}$^{*}$&\emph{9.58}$^{*}$&...&...\\
SV Cep   &$\surd$&11.40&11.23&10.97&10.59&10.14&\emph{9.35}$^{*}$&\emph{8.56}$^{*}$&\emph{7.74}$^{*}$&...&...\\
V1686Cyg &$\surd$&14.82&14.08&12.93&12.05&11.08&\emph{9.45}$^{4}$&\emph{7.60}$^{4}$&\emph{6.17}$^{4}$&4.57$^{4}$&3.60$^{4}$\\
R Mon  &...&12.22$^{*}$&12.46$^{*}$&11.85$^{*}$&10.96$^{*}$&10.25$^{*}$&\emph{9.24}$^{4}$&\emph{7.60}$^{4}$&\emph{5.83}$^{4}$&3.46$^{4}$&2.55$^{4}$\\
VY Mon&...&16.58&15.66&13.92&12.49&11.00&8.35&6.77&5.34$^{7}$&3.16$^{7}$&2.09$^{7}$\\
51 Oph   &$\surd$&4.78&4.82&4.83&4.72&4.61&\emph{4.90}$^{*}$&\emph{4.70}$^{*}$&\emph{4.30}$^{*}$&...&...\\
KK Oph   &...&12.86&12.58&12.06&11.56&10.82&8.77&7.19&5.78&3.98$^{4}$&2.51$^{4}$\\
T Ori&$\surd$&11.63&11.01&10.52&10.09&9.63&8.37&7.33&6.34&5.05$^{4}$&4.60$^{4}$\\
BF Ori   &$\surd$&10.16&9.83&9.65&9.48&9.31&8.92&8.45&7.81&6.60$^{4}$&5.80$^{4}$\\
CO Ori   &...&12.96&12.33&11.13&10.33&9.63&8.36&7.45&6.62&...&...\\
HK Ori   &...&11.82&11.83&11.44&11.02&10.56&9.22&8.18&7.26&5.87$^{4}$&5.10$^{4}$\\
NV Ori&$\surd$&10.36&10.21&9.78&9.48&9.22&8.71&8.22&7.64&...&...\\
RY Ori   &...&12.62&12.18&11.36&10.82&10.34&9.51&8.90&8.29&...&...\\
UX Ori   &$\surd$&10.25&10.04&9.80&9.62&9.43&8.92&8.31&7.62&6.10$^{4}$&5.51$^{4}$\\
V346 Ori &...&10.55&10.42&10.16&9.99&9.80&9.50&9.03&8.55&6.80$^{8}$&...\\
V350 Ori &...&12.74&12.27&11.76&11.38&10.96&9.88&9.13&8.28&...&...\\
XY Per   &...&9.91&9.47&9.05&8.70&8.40&7.59&6.83&5.99&...&...\\
VV Ser   &$\surd$&12.99&12.63&11.82&11.01&10.24&8.72&7.48&6.31&4.51$^{4}$&4.40$^{4}$\\
CQ Tau   &$\surd$&9.75&9.50&8.98&8.63&8.33&7.76&7.09&6.31&...&...\\
RR Tau   &$\surd$&11.64&11.42&10.92&10.58&10.17&8.99&7.97&7.07&5.80$^{4}$&5.50$^{4}$\\
RY Tau   &...&11.87&11.48&10.47&9.67&8.87&7.23&6.22&5.34&4.29$^{9}$&3.38$^{9}$\\
PX Vul   &...&12.56&12.29&11.51&10.90&10.30&9.31&8.52&7.77&...&...\\
WW Vul   &$\surd$&11.46&11.02&10.64&10.38&10.12&9.24&8.40&7.46&...&...\\
LkHa 234 &$\surd$&13.65&13.61&12.73&11.98&11.20&\emph{9.50}$^{10}$&\emph{8.18}$^{10}$&\emph{7.09}$^{10}$&...&...\\
\hline
\hline
\end{tabular}
\begin{minipage}{15cm}

  \underline{Notes and references to Table \ref{Table:photometry}}: Numbers are in magnitudes. Most UBVRIJHK photometry is taken from the simultaneous EXPORT multi-epoch observations \citep{Oudmaijer01,Eiroa02}, selecting the data at the brightest V-magnitude (italic numbers for the magnitudes not measured simultaneously to V). $^{*}$SIMBAD (http://simbad.u-strasbg.fr/simbad/), $^{1}$\citet{Malfait98}, $^{2}$\citet{Coulson98}, $^{3}$\citet{Sylvester96} (includes L' magnitudes for \object{HD 141569}, \object{HD 142666} and \object{HD 144432}), $^{4}$\citet{HillenbrandStrom92}, $^{5}$\citet{Bigay70}, $^{6}$\citet{Merin04}, $^{7}$\citet{Herbst82}, $^{8}$\citet{GlassPenston74}, $^{9}$\citet{KenyonHartmann95}, $^{10}$\citet{Monnier09}. Good quality IUE spectra are available for the stars indicated in Col. 2, which were used to derive synthetic UV photometry to fit the SEDs. Uncertainties are available in the references, and are typically $\lesssim$ 0.05 magnitudes for the EXPORT data.  
\end{minipage}
\end{table}
\clearpage

\begin{table}
\caption{Mid-infrared photometry}
\label{Table:photometrylong}
\centering
\begin{tabular}{lrrrrrrrrrlr}
\hline\hline
Star & ISO & F$_{9}$ & F$_{12}$ & F$_{18}$ & F$_{25}$ & F$_{60}$ & F$_{65}$ & F$_{90}$ & F$_{100}$ & F$_{140}$(F$_{160}$) & Add. Refs\\
\hline
HD 31648 &...&9.118&12.25&8.070&10.34&11.11&10.17&11.73&12.52&11.85&(1)\\
HD 34282 &...&0.736&0.700&...&1.630&10.40&10.04&...&10.72&9.745(9.353)&...\\
HD 34700 &$\surd$&...&0.600&...&4.420&14.06&...&...&9.380&...&...\\
HD 58647 &...&5.068&4.950&2.774&2.870&0.470&...&...&$<$7.360&...&...\\
HD 141569&$\surd$&0.518&0.550&0.866&1.870&5.530&6.52&3.917&3.470&3.915&...\\
HD 142666&$\surd$&5.152&8.570&6.578&11.21&7.230&5.259&5.729&5.460&5.966&...\\
HD 144432&$\surd$&...&7.530&7.326&9.360&5.760&5.290&4.916&3.290&...&(2)\\
HD150193 &$\surd$&13.00&17.61&...&18.10&8.130&5.839&6.019&$<$16.25&...&(1,2)\\
HD163296 &$\surd$&15.02&18.20&...&20.99&28.24&...&....&$<$40.62&...&(1,2,3)\\
HD 179218&$\surd$&14.74&23.44&27.88&43.63&29.92&19.43&16.80&17.35&9.263&(1)\\
HD 190073&...&5.734&7.160&4.457&5.530&1.920&...&1.119&$<$1.000&...&(1)\\
AS 442& ...&...&$<$5.188&...&2.690&$<$9.620&...&...&$<$7.215&...&...\\
VX Cas   &...&0.7554&1.540&1.327&2.640&1.400&...&0.7453&$<$15.17&...&...\\
BH Cep&...&0.481&0.475&0.767&0.116&0.138&...&0.117&$<$0.257&...&...\\
BO Cep&...&0.084&$<$0.070&0.118&0.285&1.430&...&1.566&$<$3.770&...&...\\
SV Cep   &$\surd$&2.527&4.220&3.459&5.220&2.660&...&1.858&1.760&...&...\\
V1686Cyg &$\surd$&38.73&72.80&79.53&119.0&474.0&...&...&880.0&...&(2,4)\\
R Mon    &...&34.07&54.65&79.24&132.1&121.3&129.5&78.13&148.5&66.37(62.26)&(1,2)\\
VY Mon&...&...&42.20&56.22&78.50&133.0&129.0&115.1&260.0&100.1&(4,5)\\
51 Oph   &$\surd$&...&15.67&9.241&10.19&1.060&...&...&$<$5.970&...&(1)\\
KK Oph   &...&9.100&9.870&8.276&9.560&6.140&3.905&4.431&$<$17.75&...&(2)\\
T Ori &...&...&...&...&...&...&...&...&...&...&(2)\\
BF Ori   &$\surd$&...&0.960&...&0.800&$<$2.460&...&....&$<$29.81&...&(2)\\
CO Ori   &...&1.677&1.580&1.537&1.400&$<$2.030&...&...&$<$10.88&...&...\\
HK Ori   &...&2.708&3.800&2.882&4.080&$<$1.640&...&...&$<$70.37&...&(2)\\
NV Ori & ...&1.247&...&...&...&...&...&...&...&...&...\\
RY Ori   &...&0.665&0.750&0.788&0.710&$<$2.280&...&...&$<$19.09&...&...\\
UX Ori   &$\surd$&...&2.680&3.112&3.690&2.850&2.736&2.129&3.760&4.282&(2,6)\\
V346 Ori &...&...&0.310&0.590&1.250&5.630&4.054&3.793&3.470&3.917&(7)\\
V350 Ori &...&0.556&0.940&1.008&1.750&1.190&...&...&$<$21.94&...&...\\
XY Per   &...&4.302&3.850&3.404&4.060&4.910&3.958&4.367&$<$10.27&5.323&...\\
VV Ser   &$\surd$&4.261&4.610&3.259&3.770&6.340&...&...&24.29&...&(2,8)\\
CQ Tau   &$\surd$&4.855&6.210&12.88&20.66&21.88&15.89&16.25&13.65&10.30&(6)\\
RR Tau   &$\surd$&1.365&1.740&1.367&2.200&4.460&...&...&36.68&...&(2)\\
RY Tau   &...&12.28&17.47&15.43&26.07&15.29&...&...&13.65&...&(2,6,9)\\
PX Vul   &...&0.433&0.550&0.569&0.860&$<$0.460&...&...&$<$5.060&...\\
WW Vul   &$\surd$&1.502&1.660&1.786&2.210&1.560&...&1.357&$<$9.590&...&(6)\\
LkHa 234 &$\surd$&5.844&14.78&18.72&78.96&687.7&...&...&1215&757.2(880.6)&(1,4)\\
\hline
\hline
\end{tabular}
\begin{minipage}{16.7cm}
\underline{Notes and references to Table \ref{Table:photometrylong}}: Fluxes are in Jy. IRAS fluxes at 12, 25, 60 and 100 $\mu$m are from SIMBAD (http://simbad.u-strasbg.fr/simbad/) by default, and from the IRAS Faint Source Reject Catalog (IPAC 1992) for \object{AS 442}, IRAS Faint Source Catalog, $|$b$|$ $>$ 10, Version 2.0 for \object{BH Cep} and \object{BO Cep}, and from \citet{Casey90} for \object{VY Mon}. AKARI fluxes at 9, 18, 65, 90, 140 and 160 $\mu$m are from AKARI/IRC All-Sky Survey Point Source Catalogue (Version 1.0) and AKARI/FIS All-Sky Survey Bright Source Catalogue (Version 1.0). Additional references are given in Col. 12 for photometry in the following wavelengths: (1): 10.7 $\mu$m from \citet{Monnier09}; (2): 10.5(N) and 20.1(Q) $\mu$m from \citet{HillenbrandStrom92}; (3): Spitzer IRAC 5.8 $\mu$m from the GLIMPSE catalogue; (4): Spitzer IRAC 3.6, 4.5, 5.8 and 8.0 $\mu$m from the ``Spitzer survey of young stellar clusters'' \citep{Gutermuth09}; (5): 10.5(N) and 20.1(Q) $\mu$m from \citet{Herbst82}; (6): ISO photometry at 6.9, 9.6, 17, and 28.2 $\mu$m from \citet{Thi01}; (7): Spitzer IRAC 5.8, 8.0 and MIPs 24 $\mu$m from \citet{Hernandez06}; (8,9): Spitzer IRAC 3.6, 4.5, 5.8, 8.0 and MIPs 24, 70 $\mu$m from the ``Spitzer survey of Serpens YSO population'' \citep{Harvey07} and from \citet{Robitaille07}, respectively. Photometry extracted from ISO data is available in \citet{Merin04} for the stars indicated in Col. 2. The typical uncertainty is  $\lesssim$ 10 $\%$ for IRAS and $\lesssim$ 5 $\%$ for AKARI data.
\end{minipage}
\end{table}
\clearpage
\begin{table}
\caption{Sub-mm and mm photometry for a sub-sample of stars}
\label{Table:photometryfar}
\centering
\begin{tabular}{lrrr}
\hline\hline
Star & F(1.3mm) & F($\lambda$) & Ref.\\
 & (mJy) & (mJy)((mm)) & \\
\hline
HD 31648 &360$\pm$24      &3300(0.45); 780(0.85); 235(1.4); 39.9(2.7); 35.2(2.8)                             &(1,2,3) \\
HD 34282 &110$\pm$10      &2700(0.45); 409(0.80); 360(0.85); 183(1.1); 24(2.6); 5.0(3.4)                     &(4,5,6) \\
HD 34700 &11.7$\pm$1.1$^*$&218(0.45); 41(0.85); 39(1.1)                                                      &(6,7,8) \\
HD 141569&9.0$\pm$4.0     &65(0.45); 11(0.85)                                                                &(5,9)   \\
HD 142666&127$\pm$9       &1090(0.45); 351(0.80); 180(1.1); $<$63(2.0)                                       &(6)      \\
HD 144432&37$\pm$3        &103(0.80); 129(0.85); 69(1.1)                                                     &(5.6.9) \\
HD150193 &45$\pm$4.5      &101(0.85)                                                                         &(5,10)     \\
HD163296 &780$\pm$31      &10660(0.35); 5560(0.45); 3380(0.60); 2650(0.75); 2320(0.80); 1920(0.85); 1020(1.1)&(11)          \\
HD 179218&71$\pm$7        &7.6(2.6)                                                                          &(12)          \\
VX Cas   &$<$6            &...                                                                               &(13)          \\
SV Cep   &7$\pm$2         &...                                                                               &(13)          \\
V1686Cyg &$<$4.8$^*$      &...                                                                               &(14)          \\
R Mon    &100$\pm$10      &1980(0.35); 1560(0.45); 217(0.80); 77(1.1); 1.17(20); 0.37(36); 0.29(60)          &(11,15,16)\\
VY Mon   &120$\pm$12      &18000(0.37); 1.5(60)                                                              &(10,17)    \\
KK Oph   &52$\pm$5.2      &109(0.80); 36(1.1); 0.24(36)                                                      &(5,16)     \\
T Ori    &130$\pm$28      &...                                                                               &(15)          \\
BF Ori   &6$\pm$2         &...                                                                               &(13)         \\
CO Ori   &$<$4.0          &...                                                                               &(13)         \\
HK Ori   &$<$100          &...                                                                               &(15)        \\
UX Ori   &23$\pm$2        &3.8(2.6); 0.8(7.0)                                                                &(13,18,19)      \\
VV Ser   &$<$20           &...                                                                               &(15)            \\
CQ Tau   &143$\pm$8.4     &2.6(7.0)                                                                          &(12,19)       \\
RR Tau   &$<$20           &...                                                                               &(15)          \\
RY Tau   &212$\pm$21$^*$  &2439(0.35); 1920(0.45); 560(0.85)                                                 &(20,21)       \\
WW Vul   &10.5$\pm$1.0    &...                                                                               &(13)          \\
LkHa 234 &827$\pm$82.7$^*$&60000(0.45); 4200(0.80); 3400(0.85); 2.67(20); 1.82(36); 1.65(60)                 &(16)          \\
\hline
\hline
\end{tabular}
\begin{minipage}{16.7cm}
Notes. The second column shows the fluxes used to compute the disk masses in Table \ref{Table:sample}. Those tagged with $^*$ were measured at
1.35 mm (\object{HD 34700}), 2.70 mm (\object{V1686 Cyg}) and 1.20 mm (\object{RY Tau} and \object{LkHa 234}). An uncertainty of 10$\%$  is assigned when that is not provided in the reference. 
(1): \citet{ManningsSargent97}, (2): \citet{SandellWein03}, (3): \citet{Pietu06},(4): \citet{Pietu03}, (5): \citet{Sandell11},
(6): \citet{Sylvester96},(7): \citet{Sylvester01}, (8): \citet{Sheret04}, (9): \citet{WalkerButner95}, (10): \citet{AlonsoAlbi09},
(11): \citet{Mannings94}, (12): \citet{ManningsSargent00}, (13): \citet{Natta97}, (14): \citet{Natta00}, (15):
\citet{HillenbrandStrom92}, (16): \citet{PezzutoStrafella97}, (17): \citet{Casey90}, (18): \citet{Acke06}, (19): \citet{Testi01},
(20): \citet{Robitaille07}, (21): \citet{Altenhoff94}
\end{minipage}
\end{table}

\clearpage

\begin{figure*}
\centering
\begin{tabular}{cc}
\includegraphics[width=80mm,clip=true]{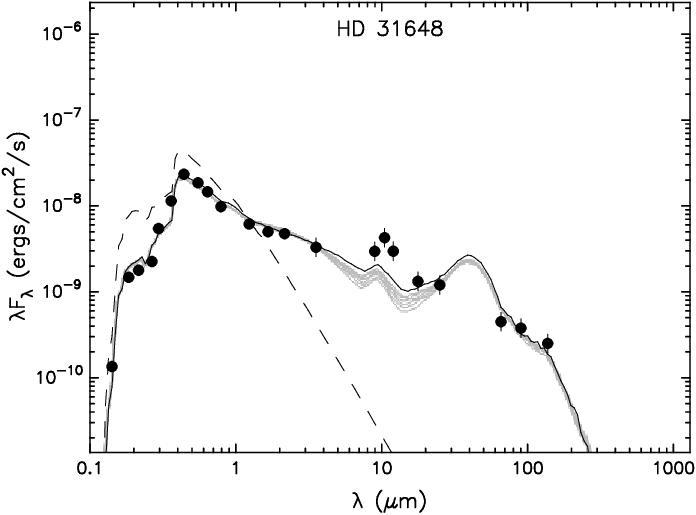} &
\includegraphics[width=80mm,clip=true]{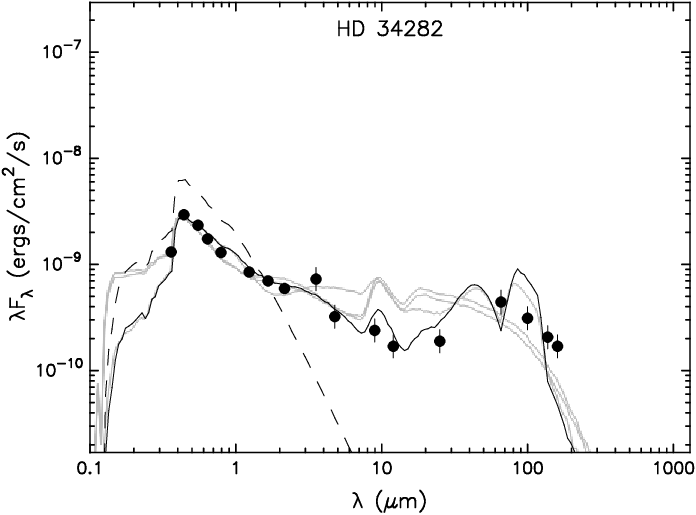} \\
\includegraphics[width=80mm,clip=true]{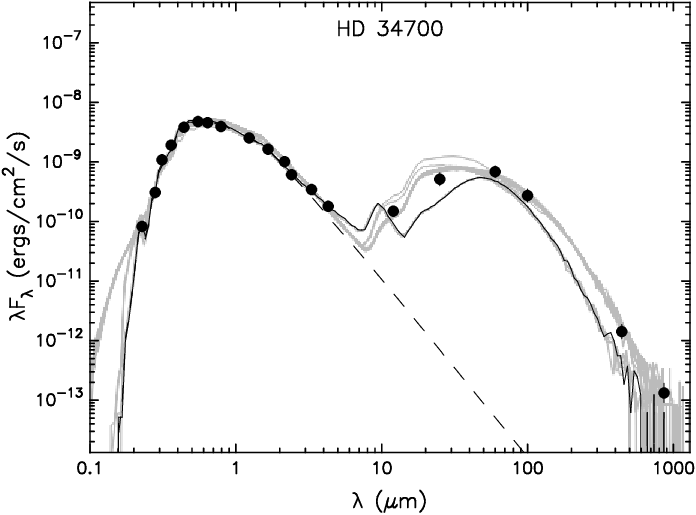} &
\includegraphics[width=80mm,clip=true]{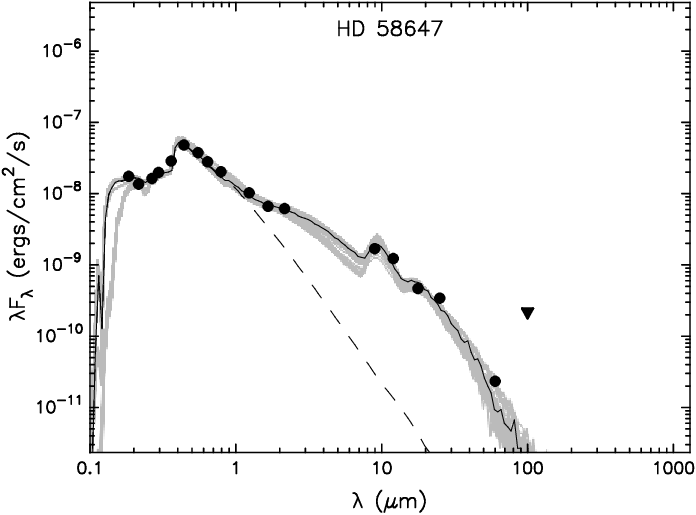}\\ 
\includegraphics[width=80mm,clip=true]{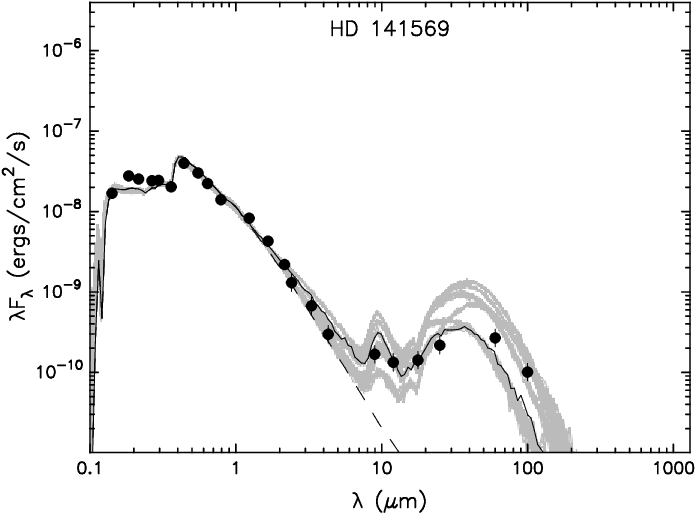} &
\includegraphics[width=80mm,clip=true]{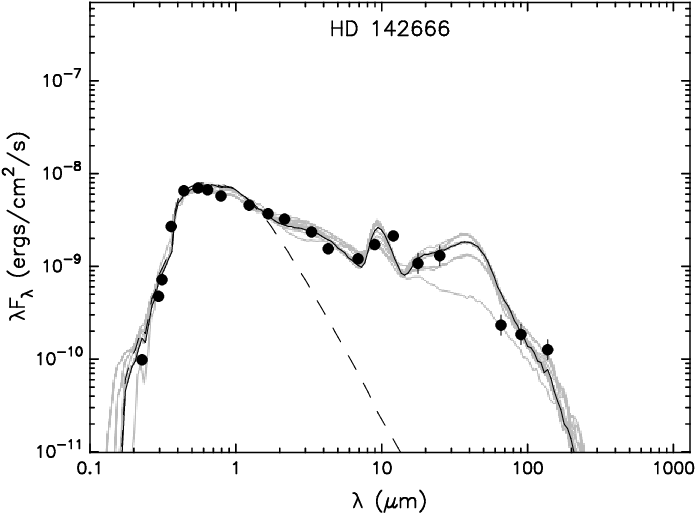} \\
\includegraphics[width=80mm,clip=true]{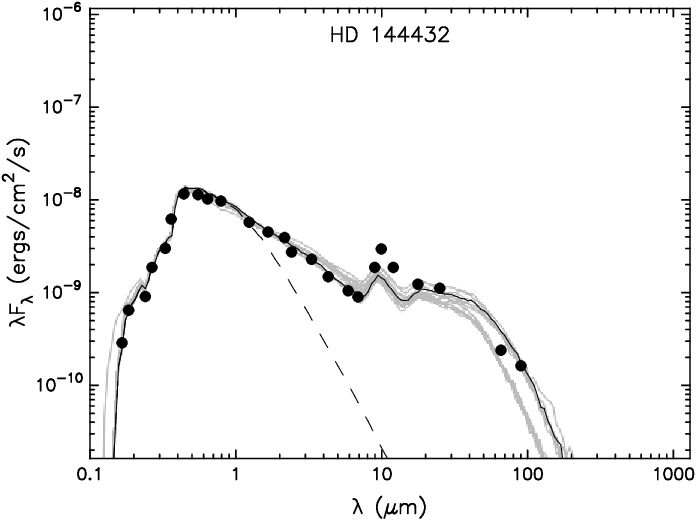} &
\includegraphics[width=80mm,clip=true]{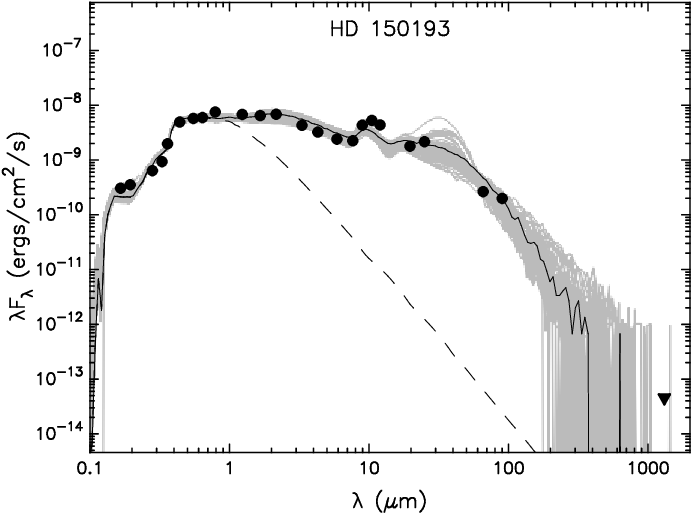} \\
\end{tabular}
\end{figure*}
\begin{figure*}
\centering
\begin{tabular}{cc}
\includegraphics[width=80mm,clip=true]{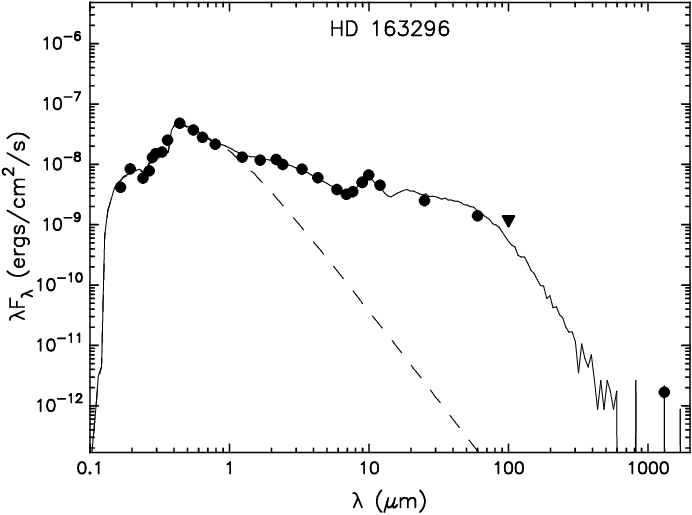} &
\includegraphics[width=80mm,clip=true]{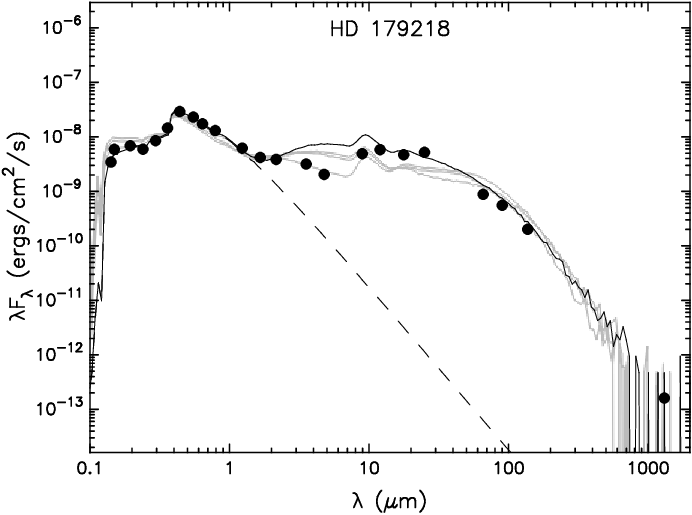} \\
\includegraphics[width=80mm,clip=true]{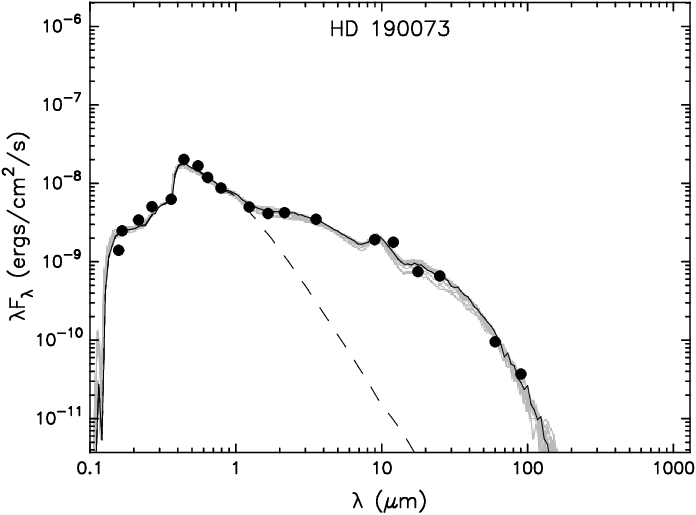} &
\includegraphics[width=80mm,clip=true]{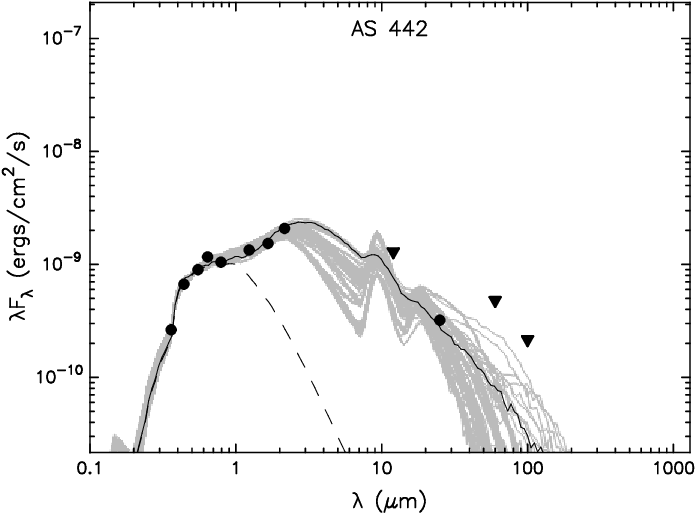}\\ 
\includegraphics[width=80mm,clip=true]{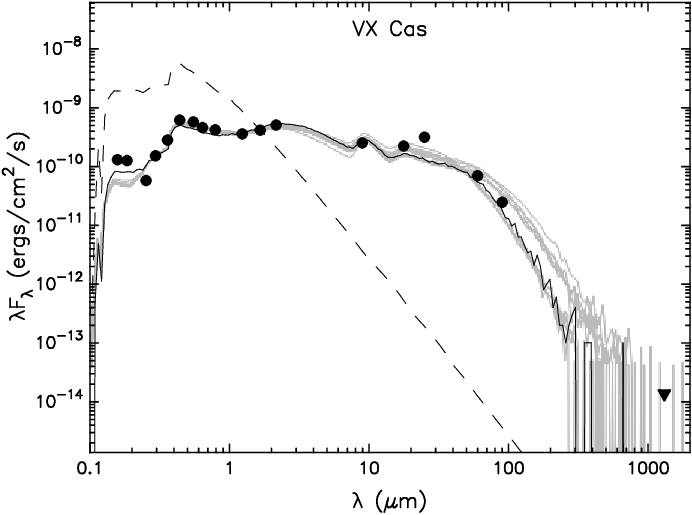} &
\includegraphics[width=80mm,clip=true]{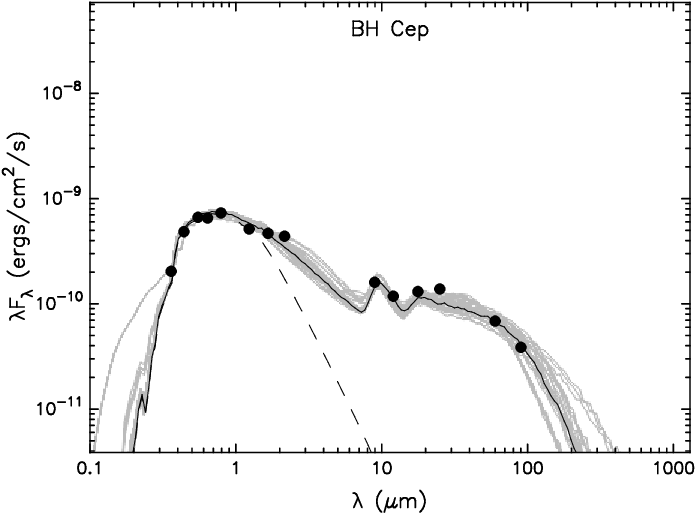}\\
\includegraphics[width=80mm,clip=true]{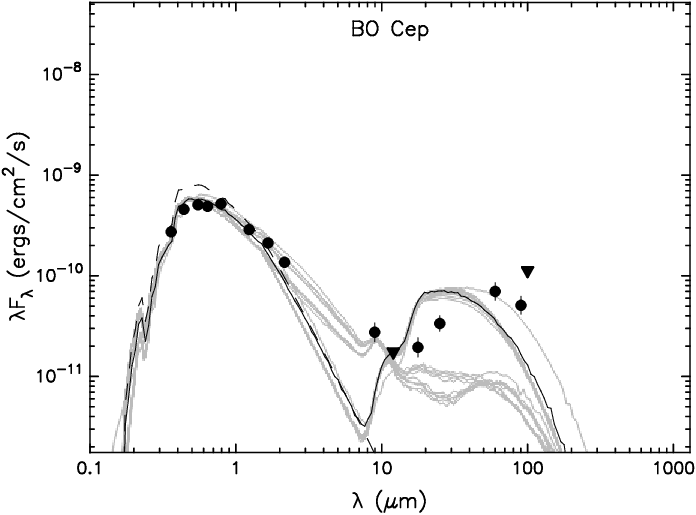}&
\includegraphics[width=80mm,clip=true]{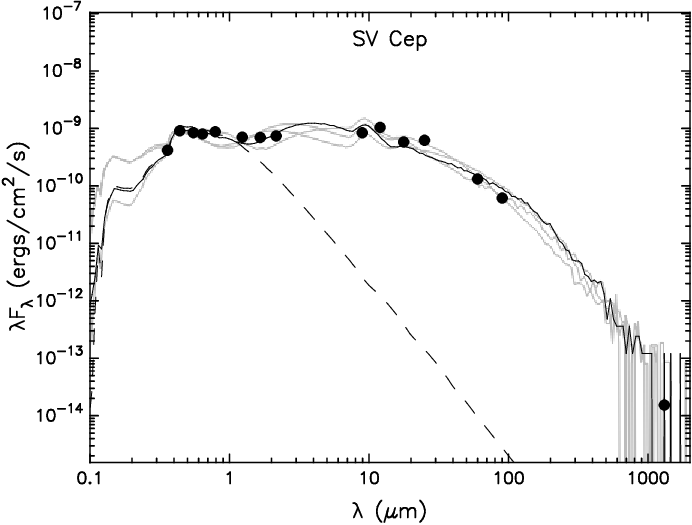} \\
\end{tabular}
\end{figure*}
\begin{figure*}
\centering
\begin{tabular}{cc}
\includegraphics[width=80mm,clip=true]{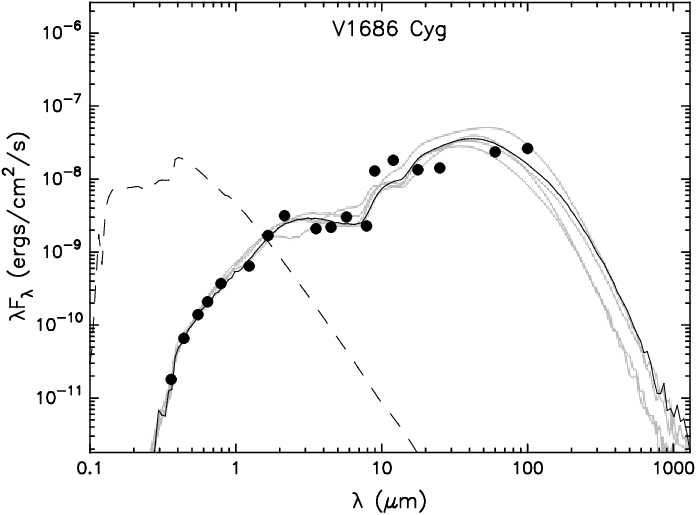} &
\includegraphics[width=80mm,clip=true]{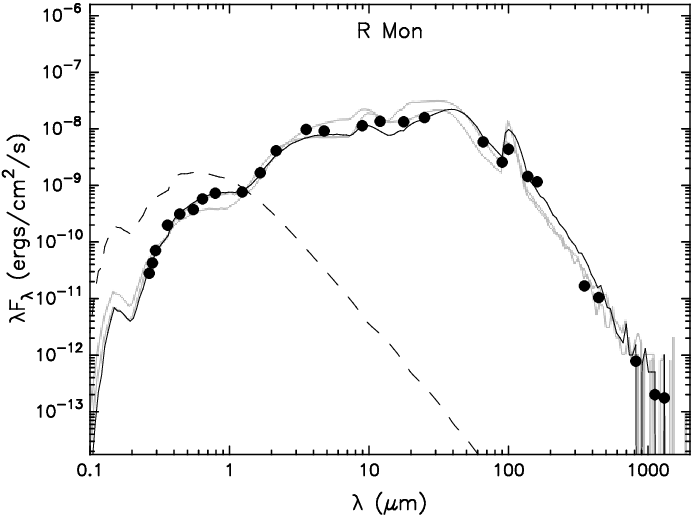} \\
\includegraphics[width=80mm,clip=true]{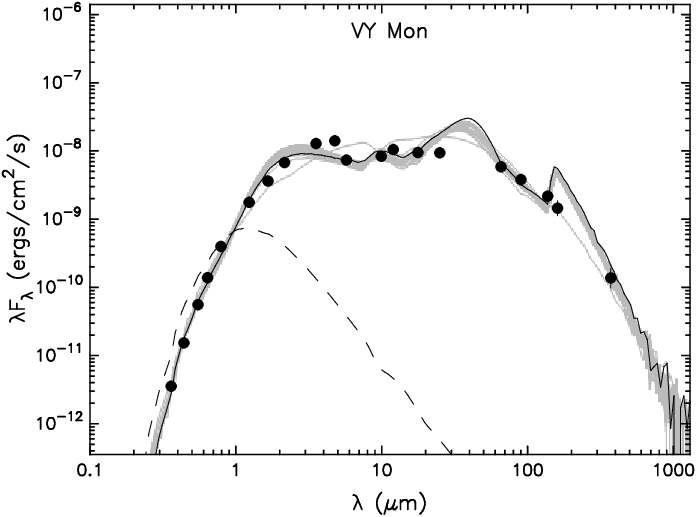}&
\includegraphics[width=80mm,clip=true]{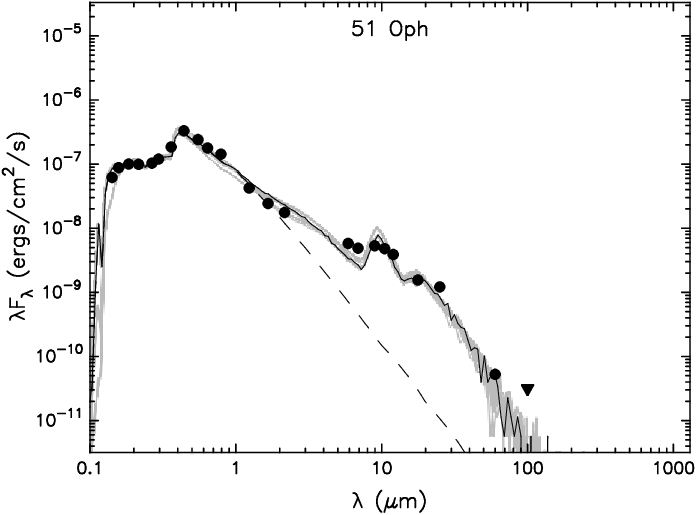} \\
\includegraphics[width=80mm,clip=true]{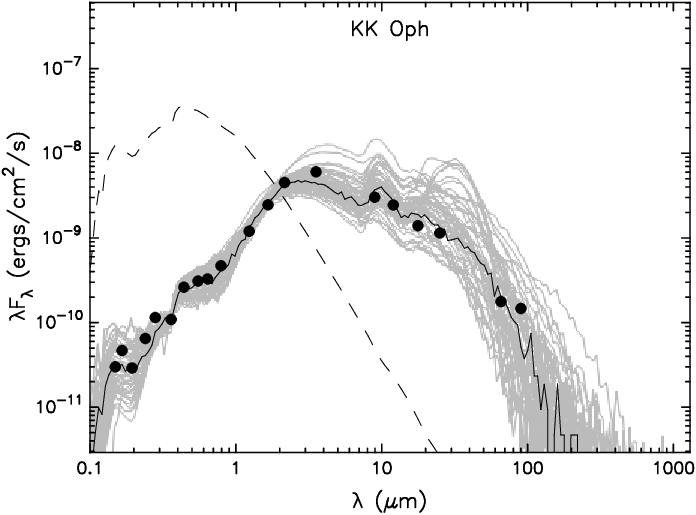} &
\includegraphics[width=80mm,clip=true]{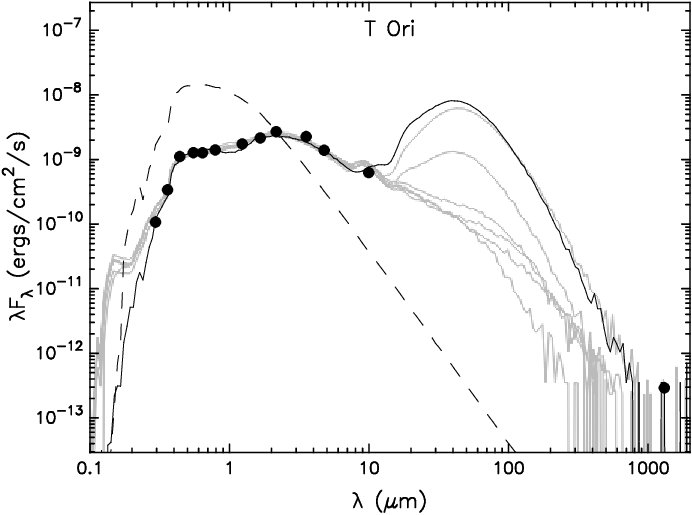}\\
\includegraphics[width=80mm,clip=true]{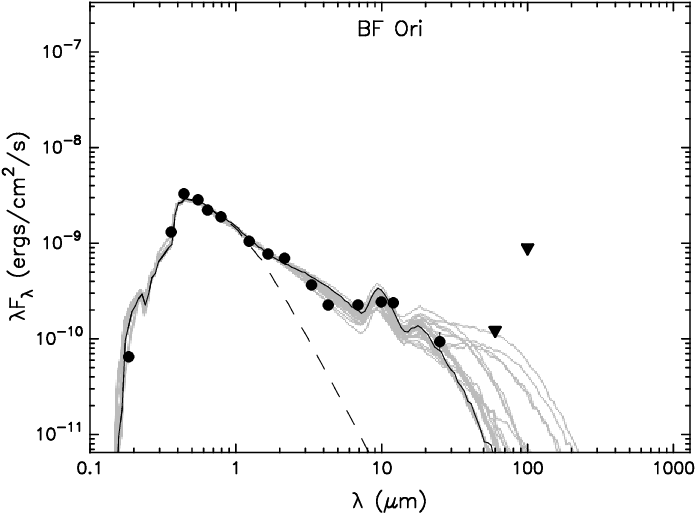} &
\includegraphics[width=80mm,clip=true]{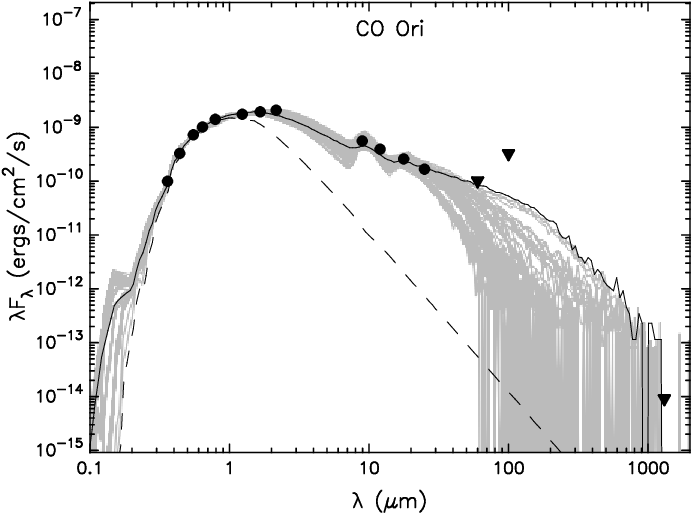} \\
\end{tabular}
\end{figure*}
\begin{figure*}
\centering
\begin{tabular}{cc}
\includegraphics[width=80mm,clip=true]{figures/hkori.eps} &
\includegraphics[width=80mm,clip=true]{figures/nvori.eps}\\
\includegraphics[width=80mm,clip=true]{figures/ryori.eps} &
\includegraphics[width=80mm,clip=true]{figures/uxori.eps} \\
\includegraphics[width=80mm,clip=true]{figures/v346ori.eps} &
\includegraphics[width=80mm,clip=true]{figures/v350ori.eps} \\
\includegraphics[width=80mm,clip=true]{figures/xyper.eps} &
\includegraphics[width=80mm,clip=true]{figures/vvser.eps} \\
\end{tabular}
\end{figure*}
\begin{figure*}
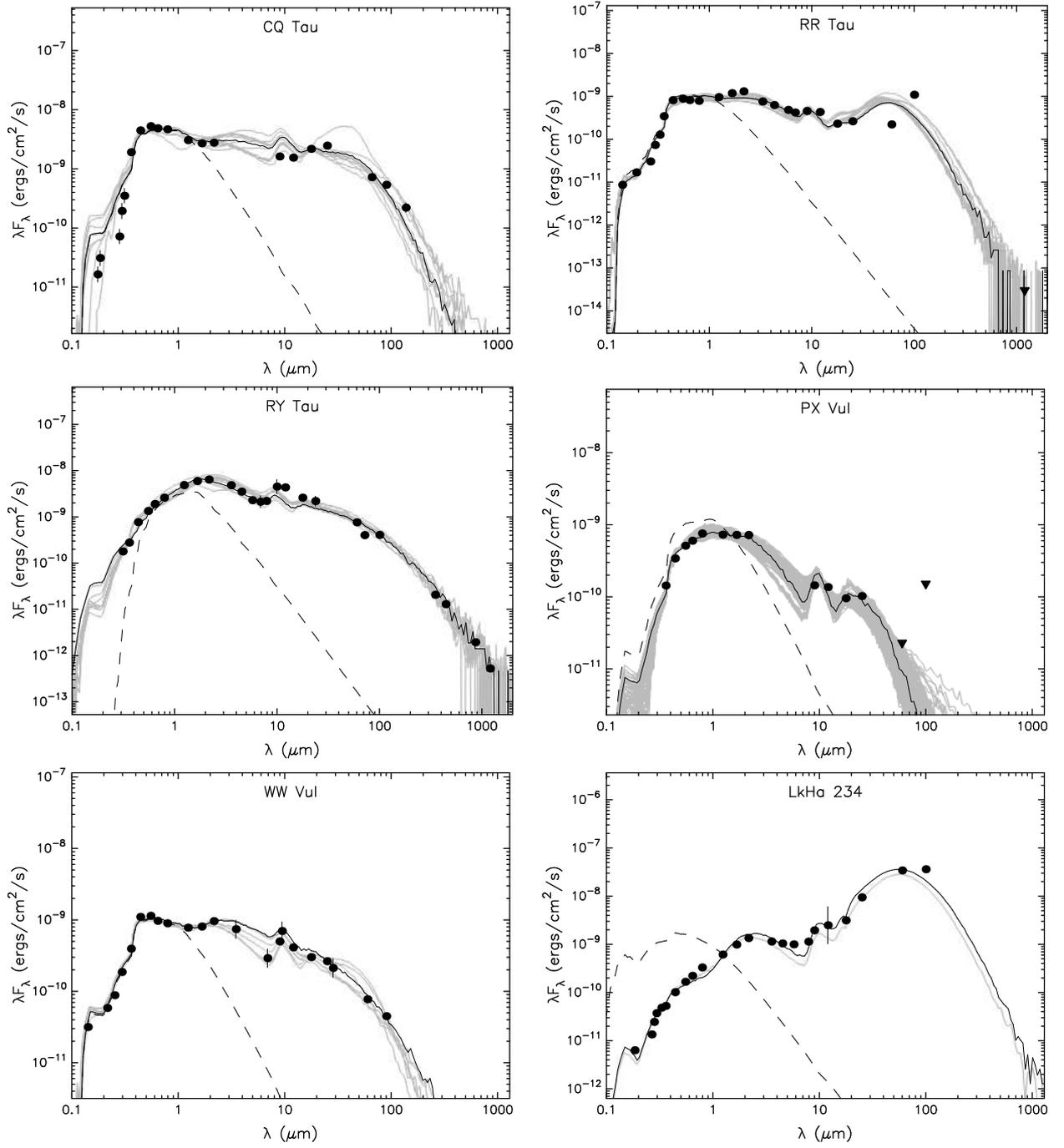

\centering
\begin{tabular}{cc}
\includegraphics[width=80mm,clip=true]{figures/cqtau.eps} &
\includegraphics[width=80mm,clip=true]{figures/rrtau.eps} \\
\includegraphics[width=80mm,clip=true]{figures/rytau.eps} &
\includegraphics[width=80mm,clip=true]{figures/pxvul.eps} \\
\includegraphics[width=80mm,clip=true]{figures/wwvul.eps} &
\includegraphics[width=80mm,clip=true]{figures/lkha234.eps}\\
\end{tabular}
\caption{SEDs of the stars in the sample. Triangles are photometric upper limits. For each panel, the best fit is the black-solid line and the models fitting with $\chi$$^{2}$ - $\chi$$^{2}_{best}$ $\leq$ 3n are the grey-solid lines, where n is the number of datapoints per star \citep[see][]{Robitaille07}. The best stellar (de-redenned) photosphere is indicated with a dashed line.}
\label{Figure:SEDs}
\end{figure*} 

\begin{figure*}
\centering
\begin{tabular}{cc}
\includegraphics[width=70mm,clip=true]{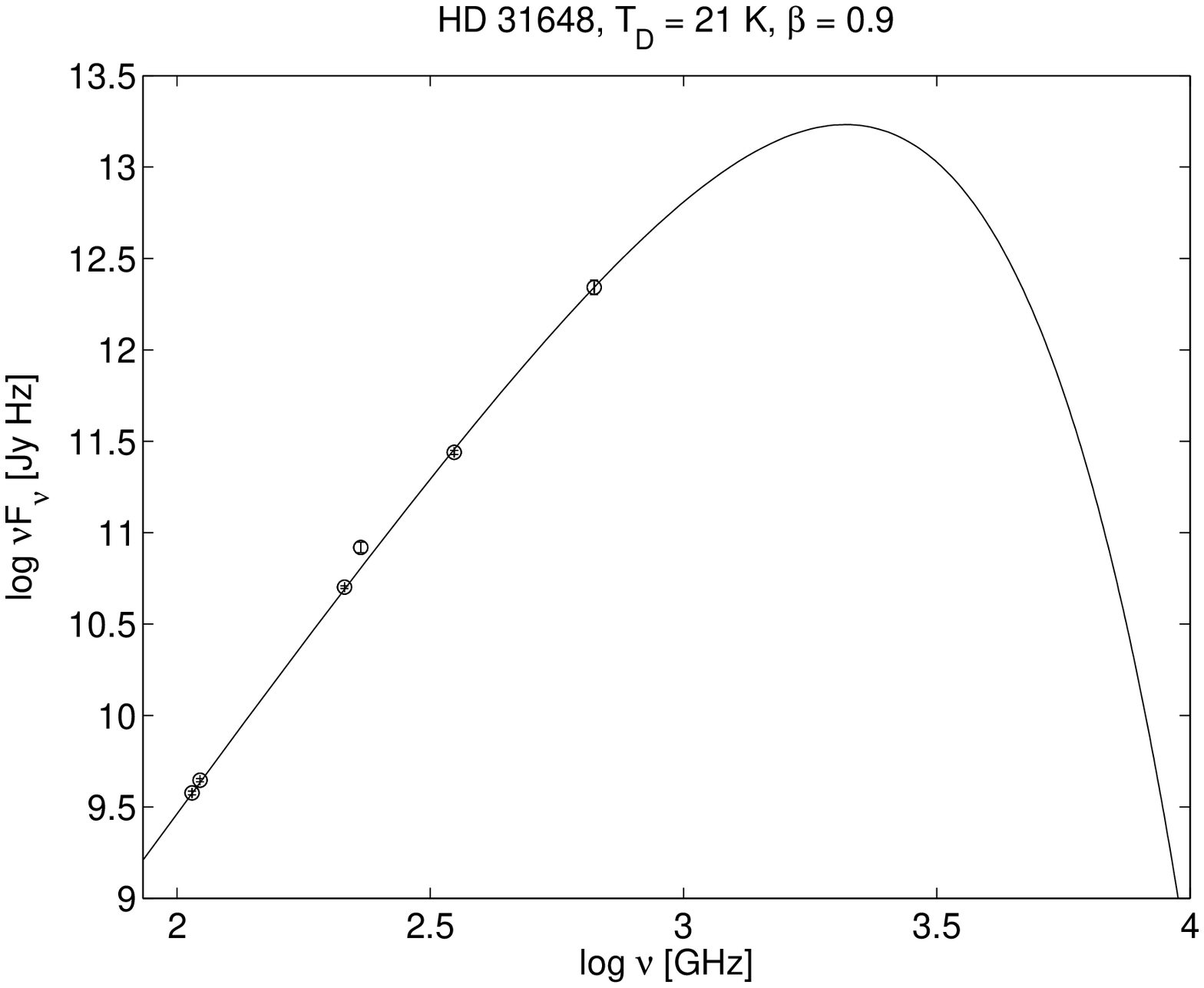} &
\includegraphics[width=70mm,clip=true]{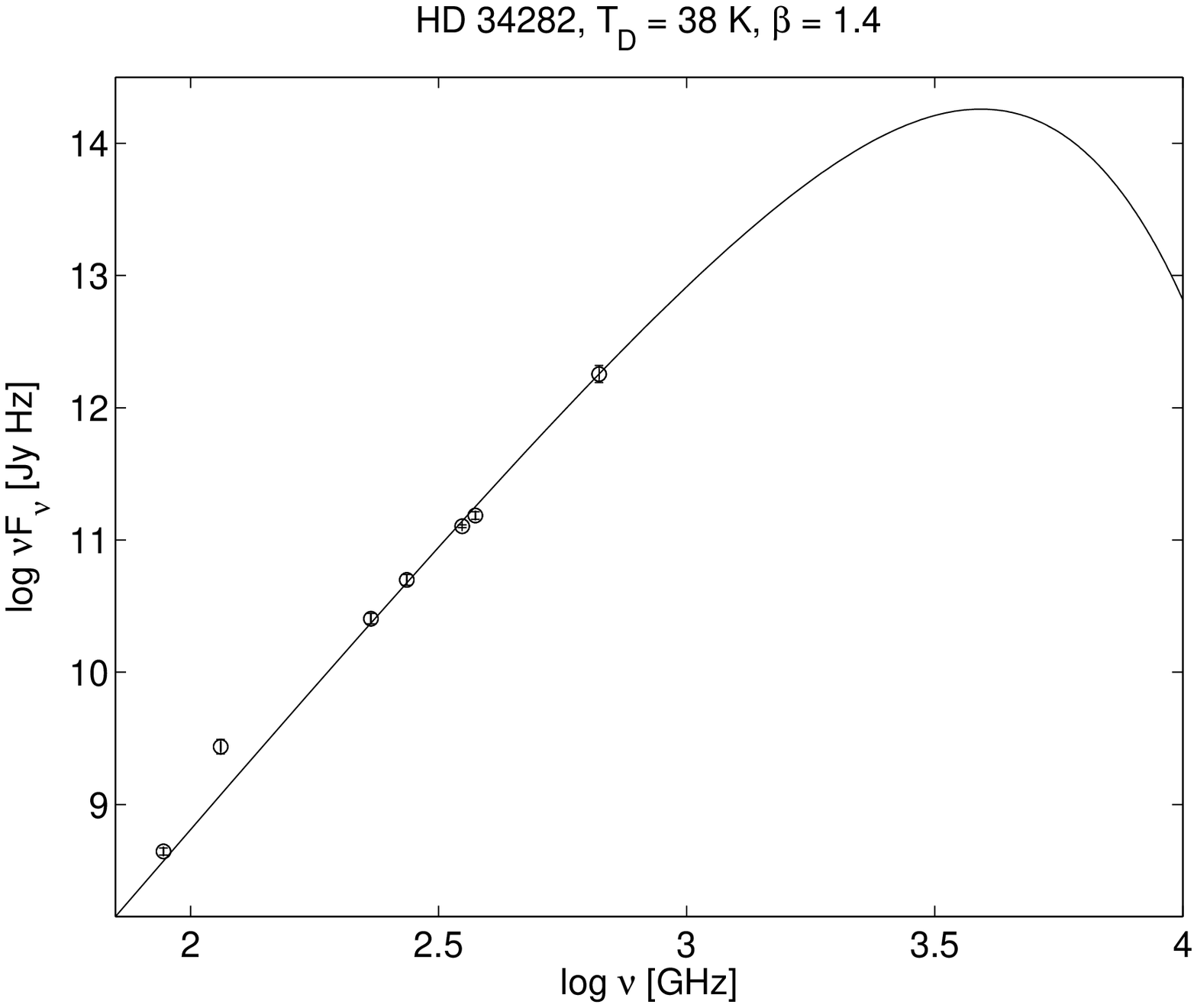} \\
\includegraphics[width=70mm,clip=true]{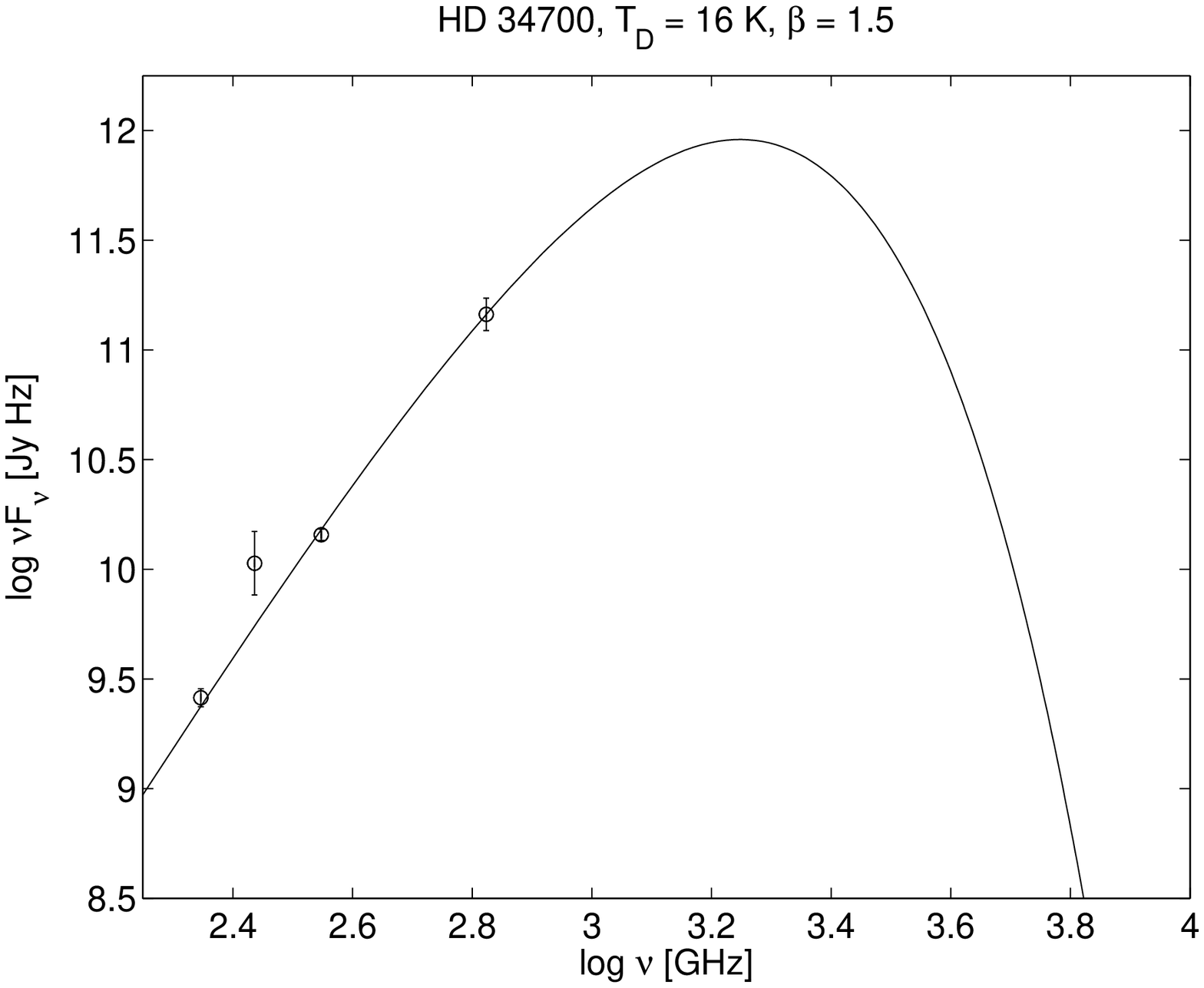} &
\includegraphics[width=70mm,clip=true]{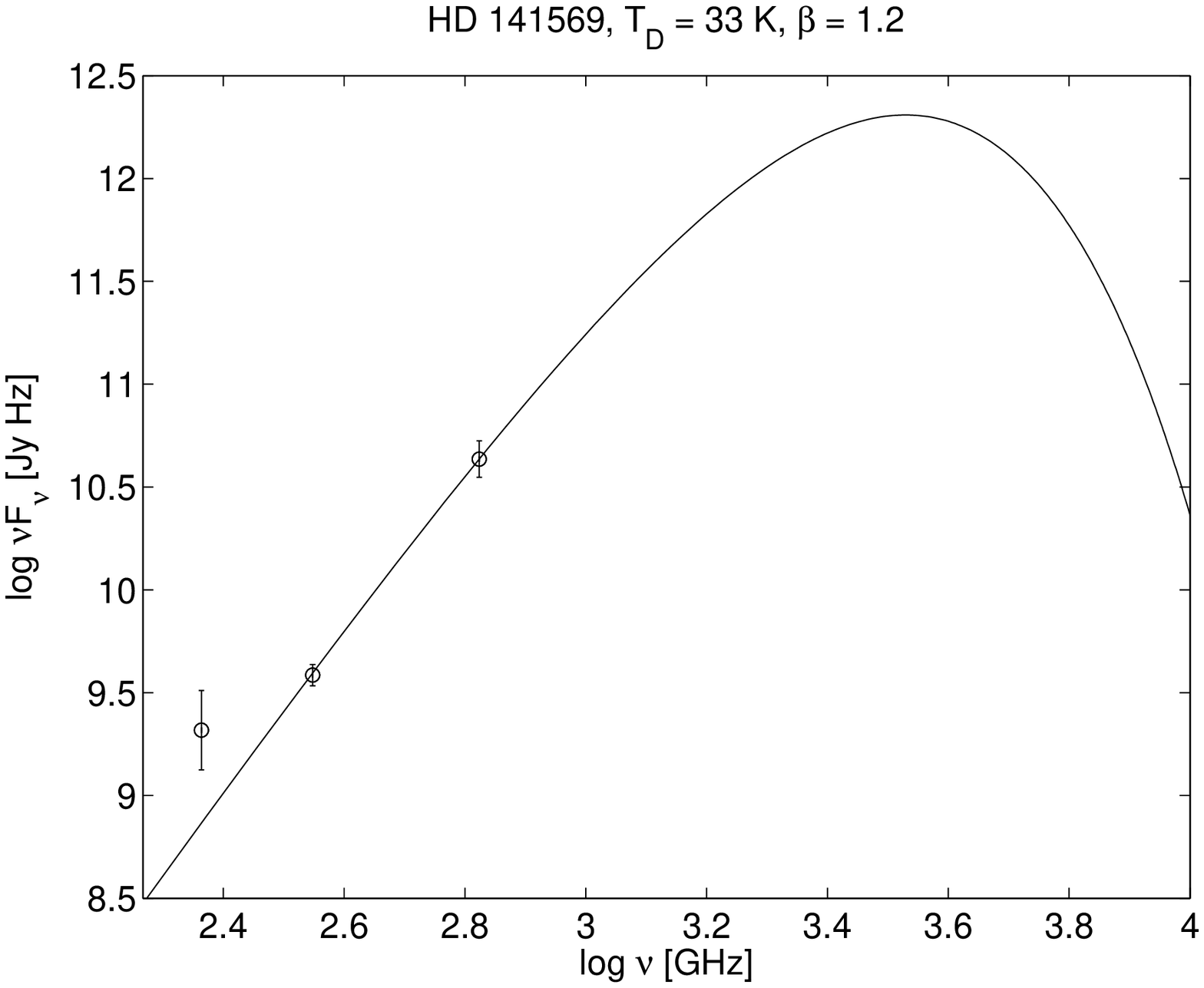}\\ 
\includegraphics[width=70mm,clip=true]{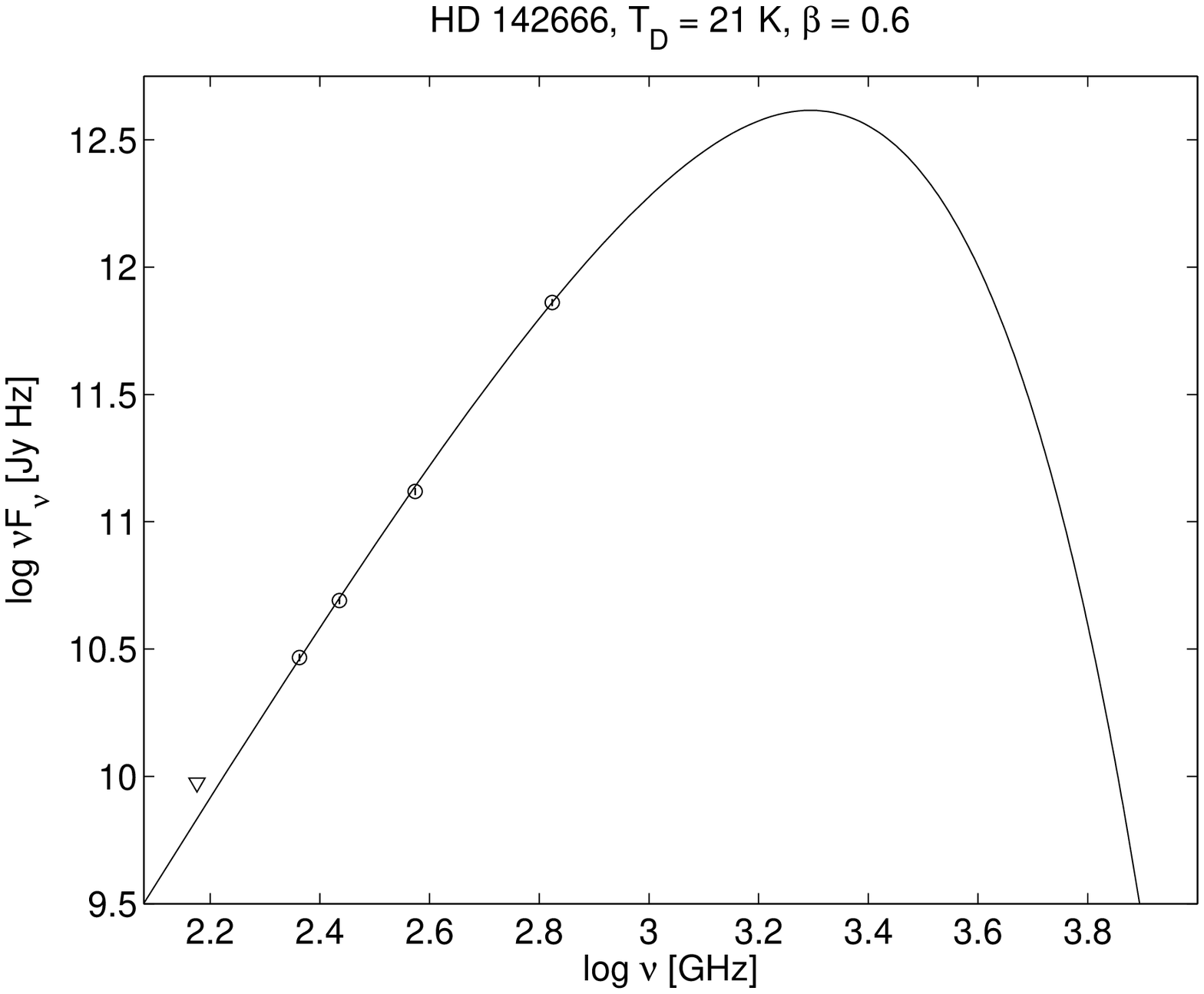} &
\includegraphics[width=70mm,clip=true]{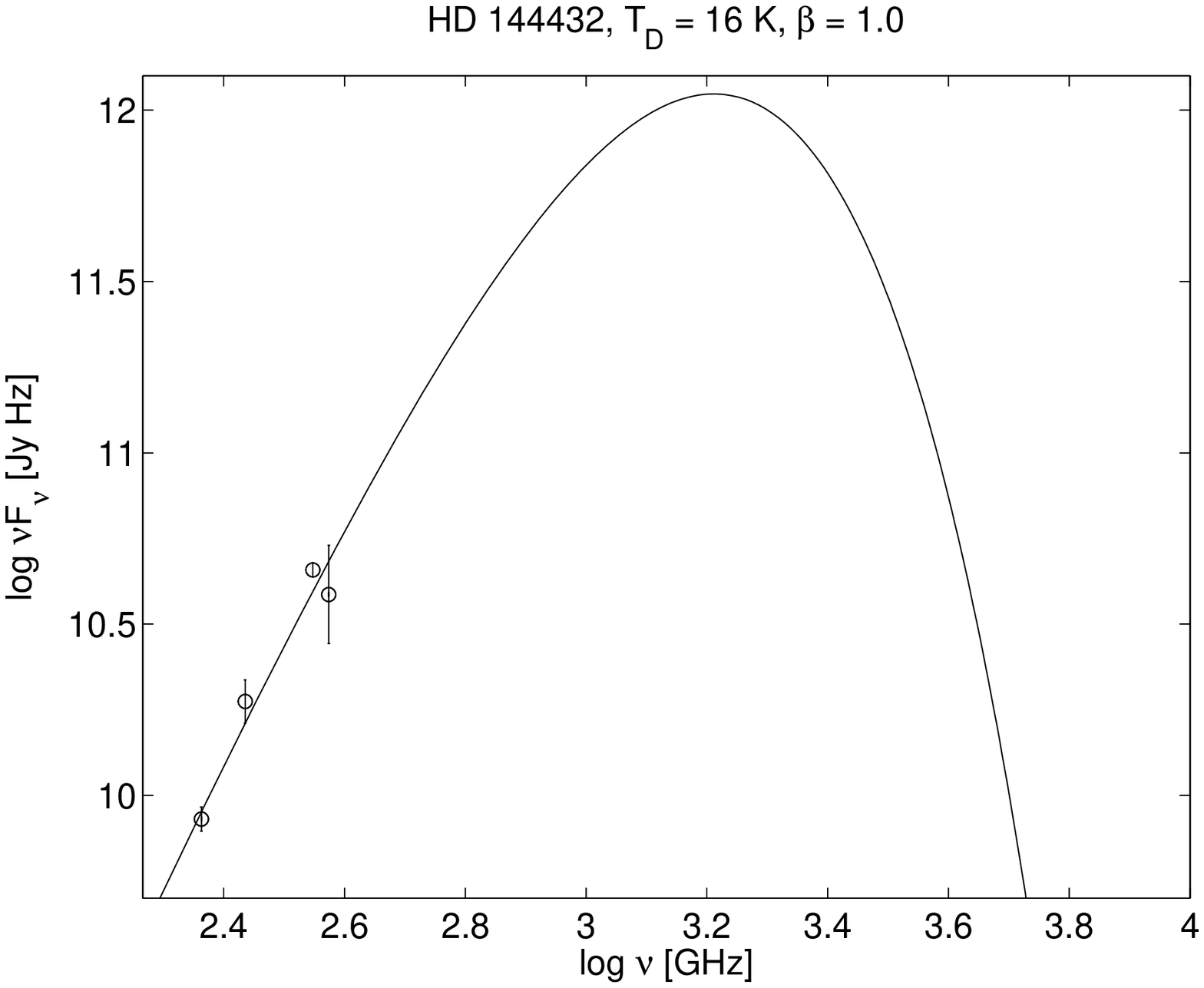} \\
\includegraphics[width=70mm,clip=true]{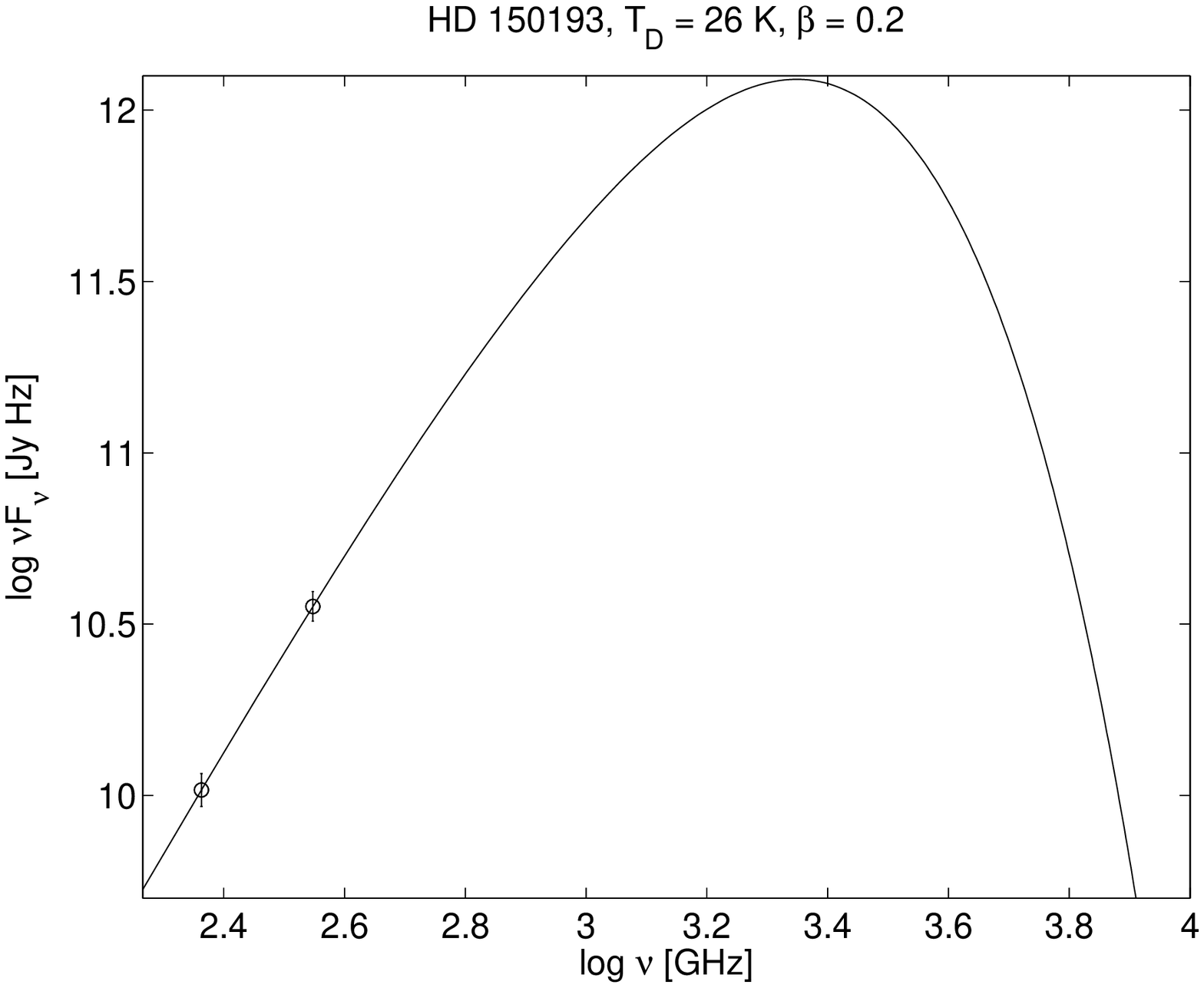} &
\includegraphics[width=70mm,clip=true]{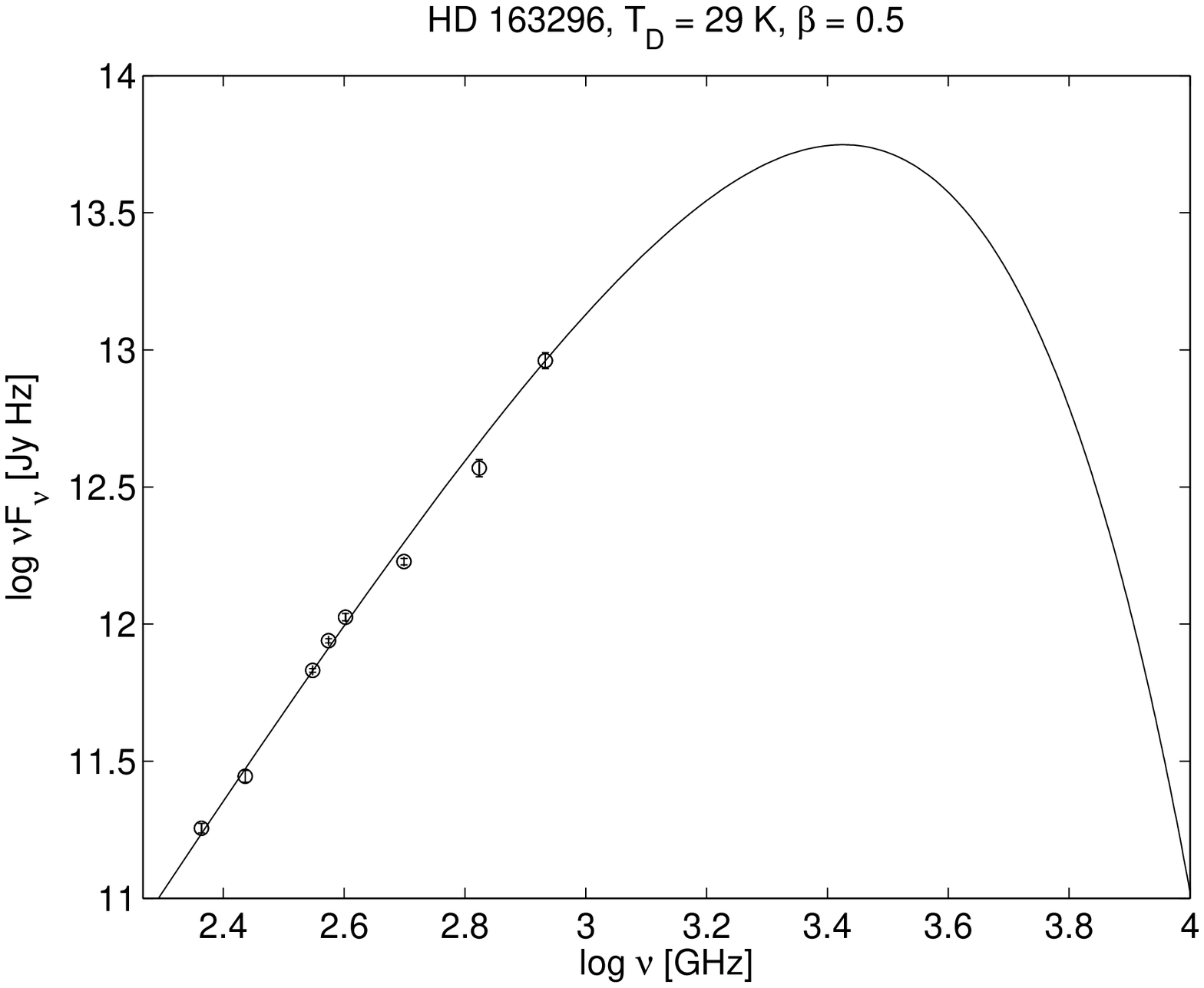} \\
\end{tabular}
\end{figure*}
\begin{figure*}
\centering
\begin{tabular}{cc}
\includegraphics[width=70mm,clip=true]{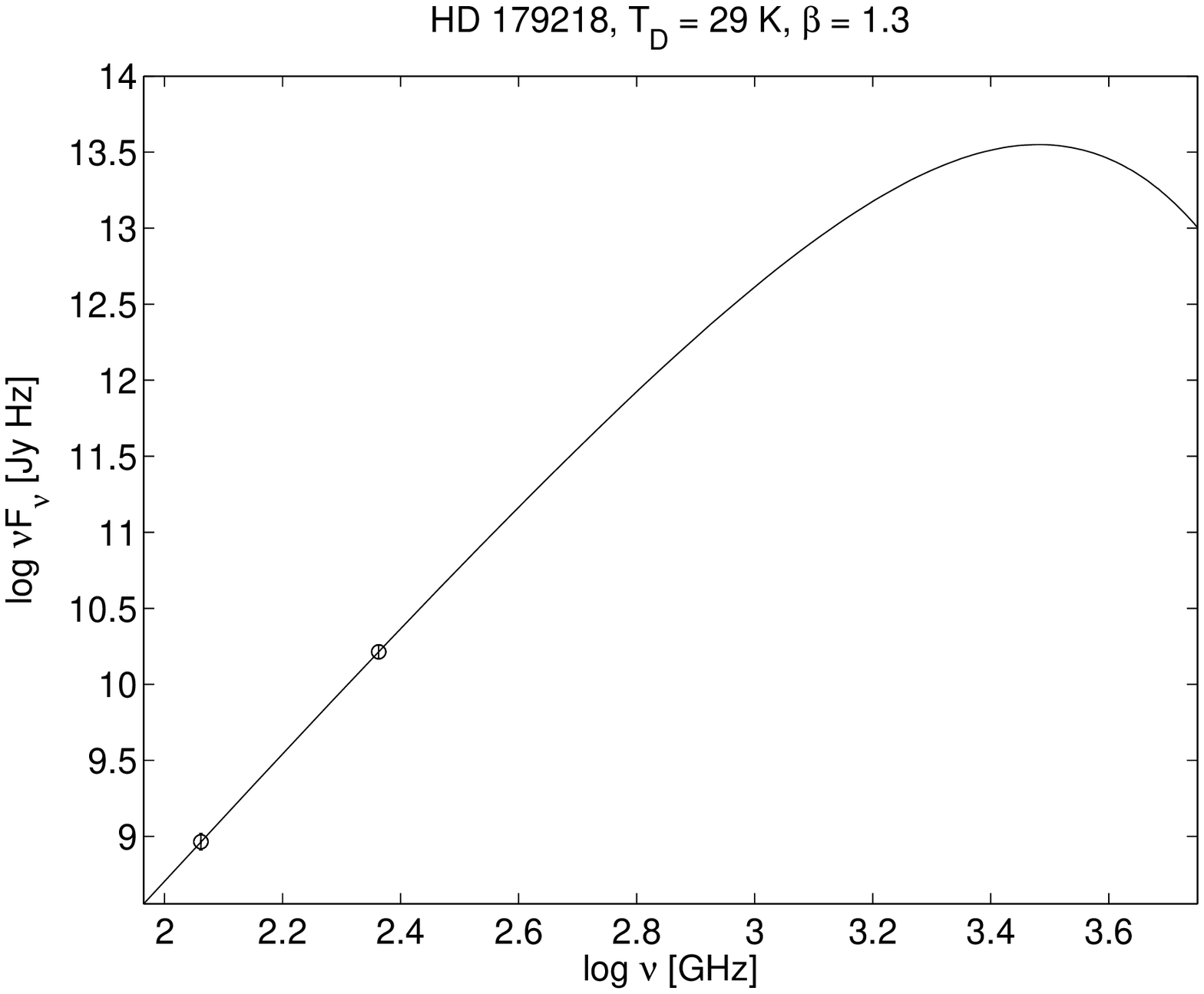} &
\includegraphics[width=70mm,clip=true]{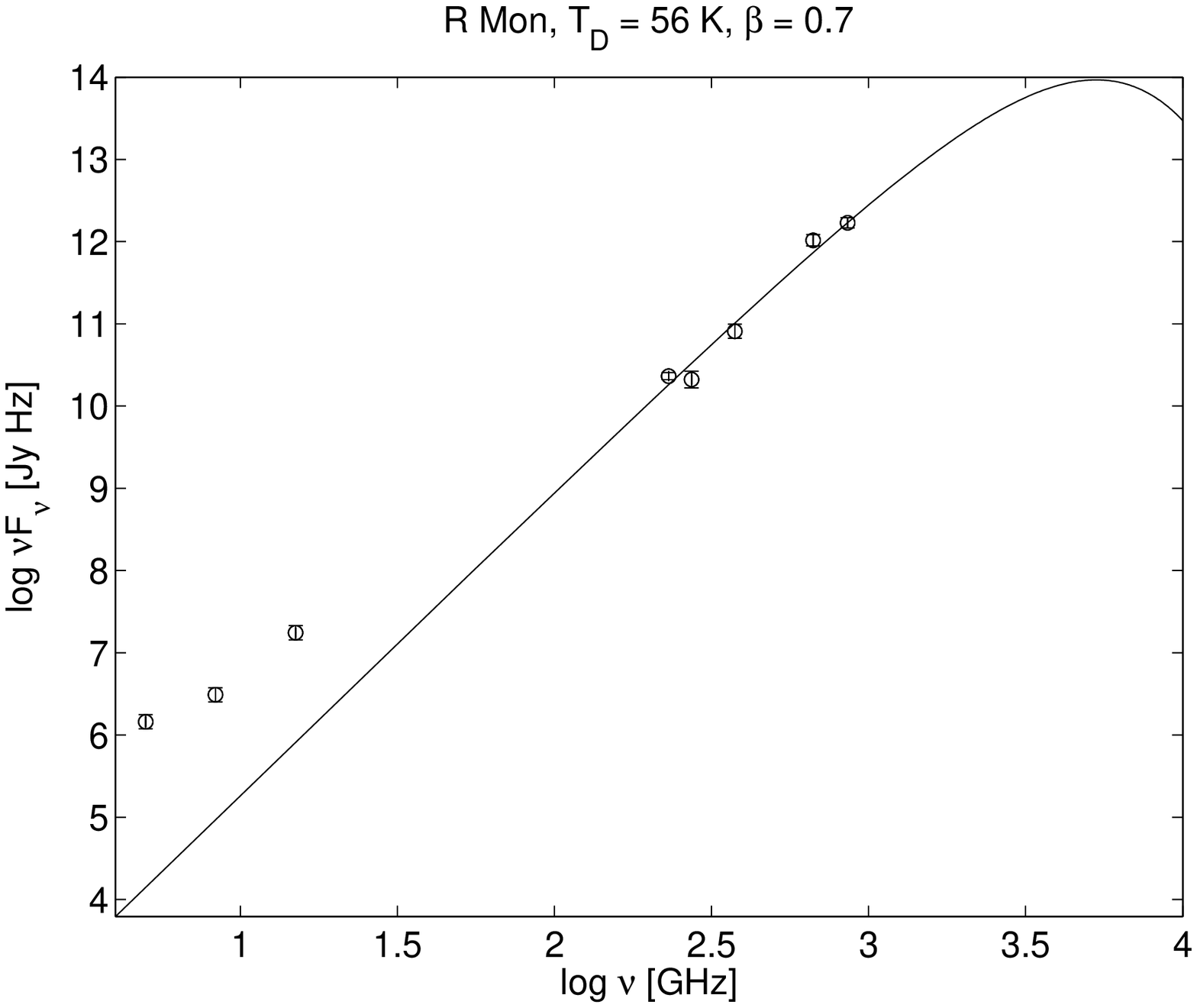} \\
\includegraphics[width=70mm,clip=true]{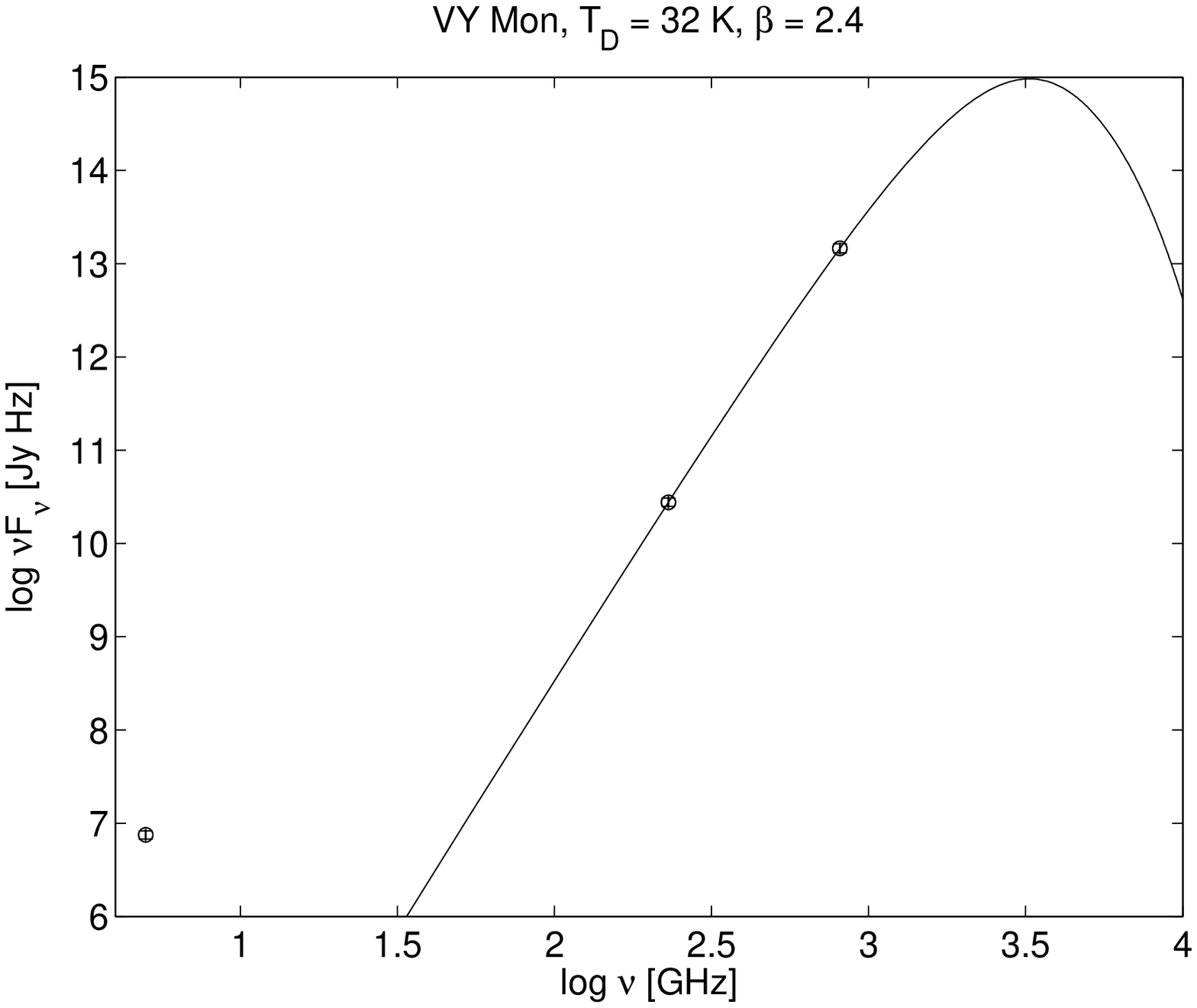} &
\includegraphics[width=70mm,clip=true]{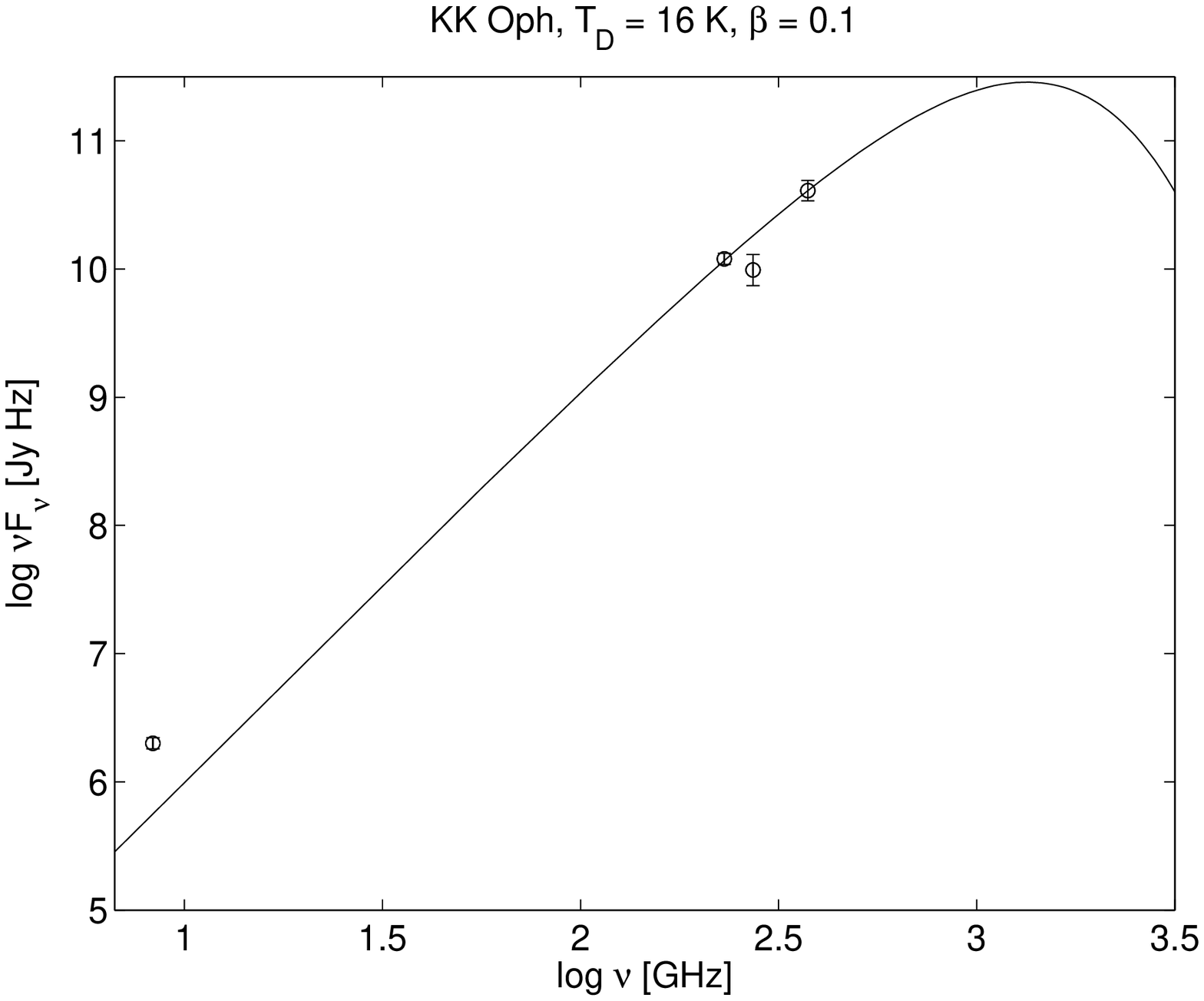}\\ 
\includegraphics[width=70mm,clip=true]{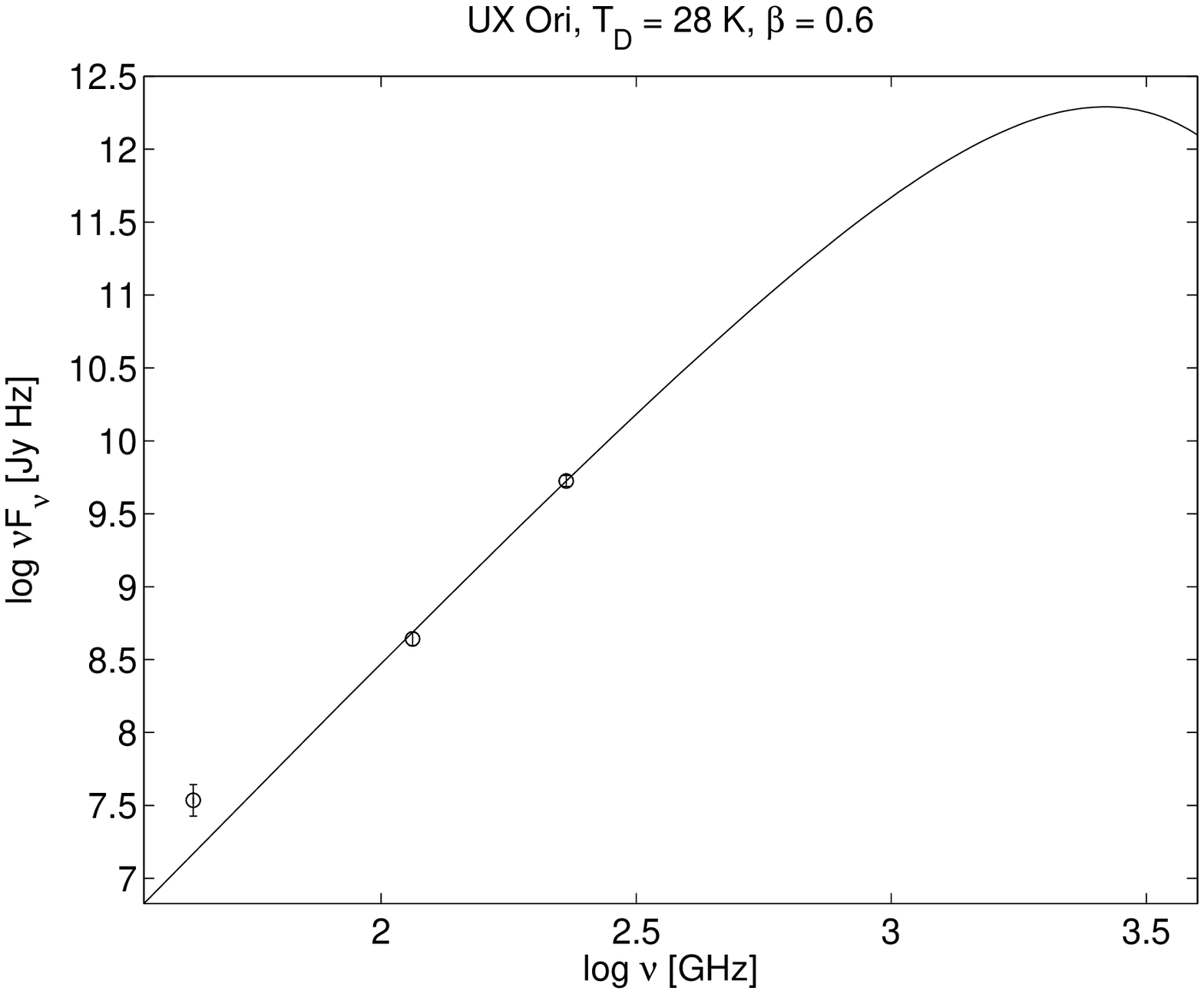} &
\includegraphics[width=70mm,clip=true]{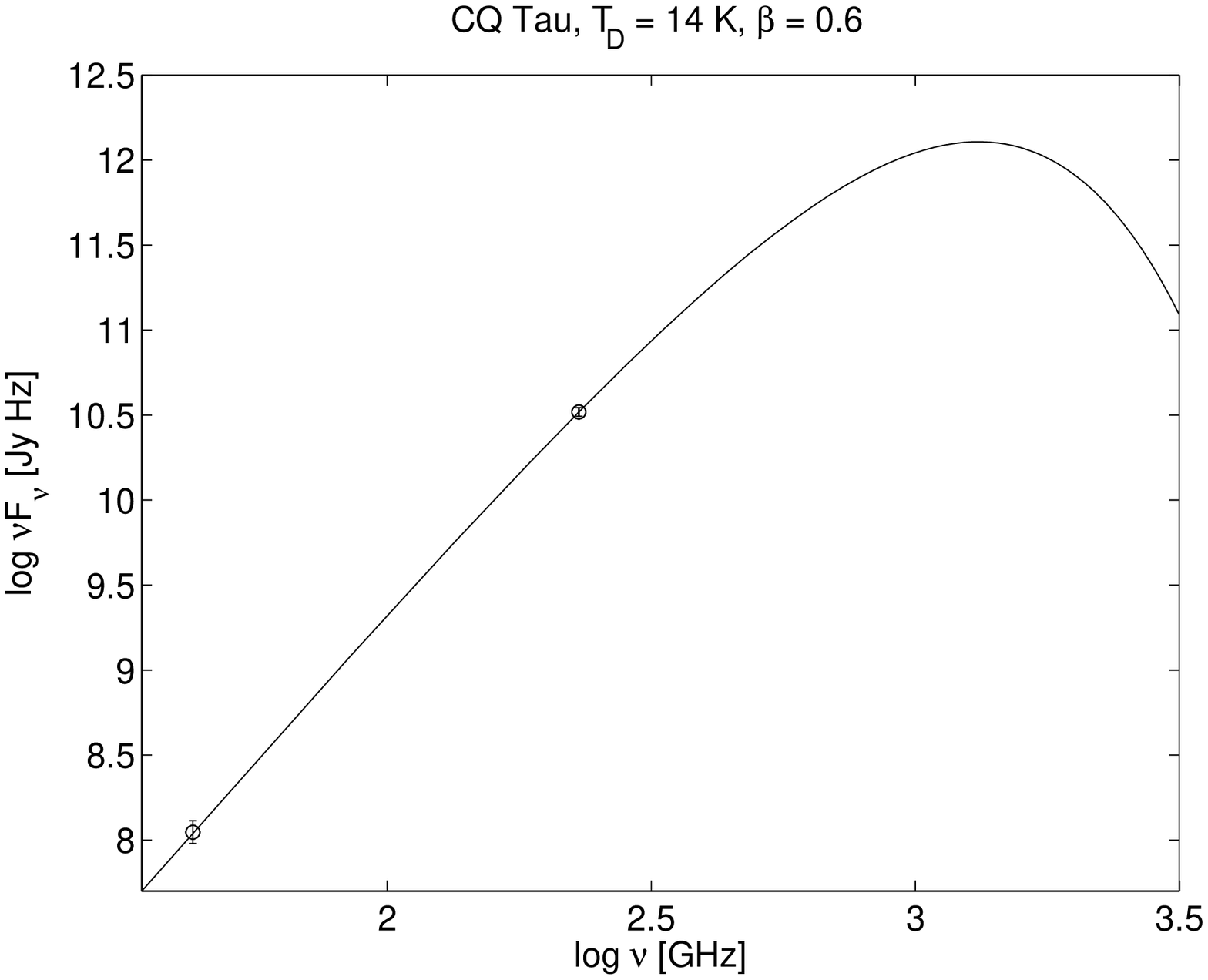}\\
\includegraphics[width=70mm,clip=true]{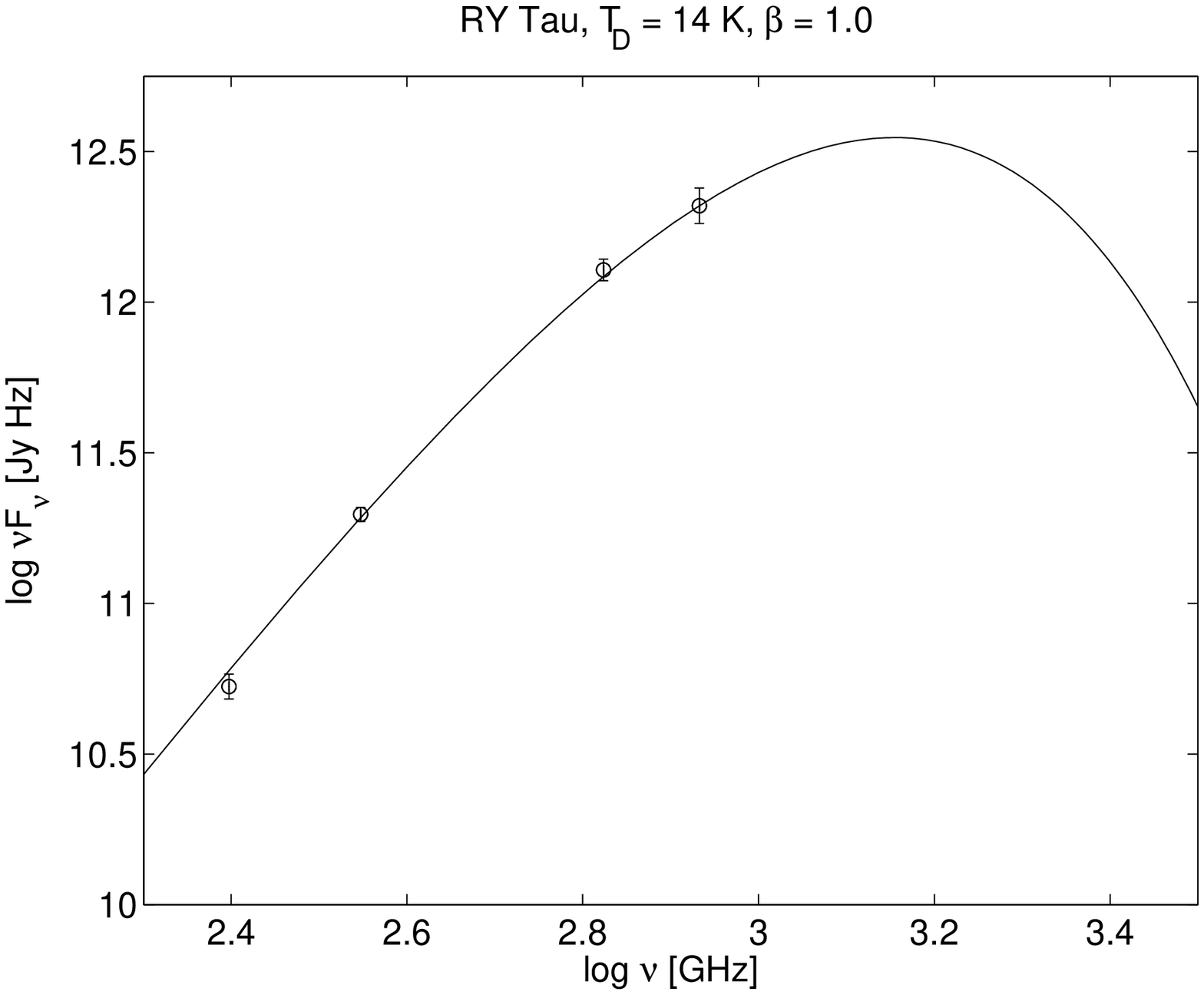}&
\includegraphics[width=70mm,clip=true]{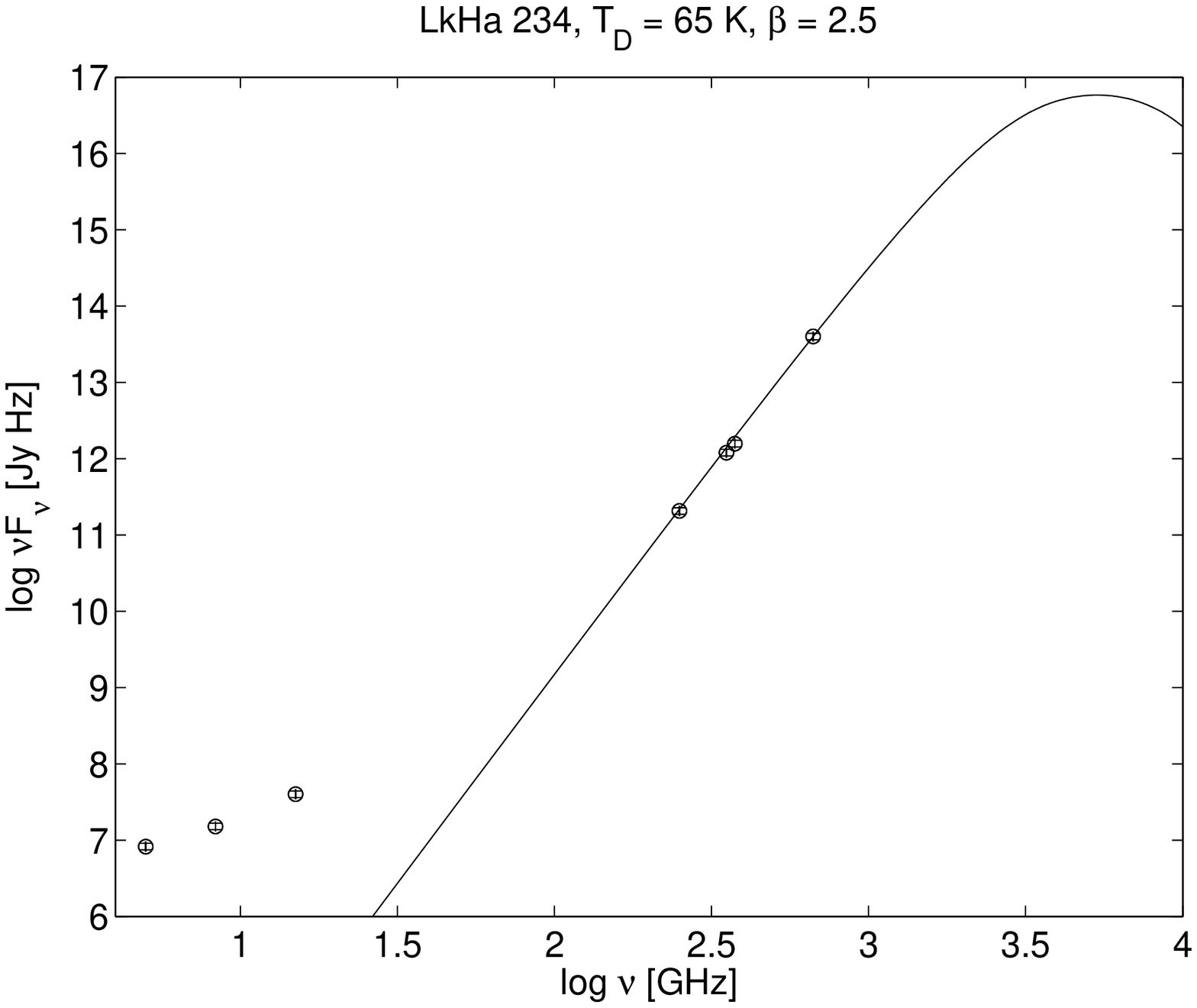} \\
\end{tabular}
\caption{Subsample of stars with fluxes available at $\lambda$ $\geq$ 0.35 mm (log $\nu$ [GHz] $\leq$ 2.9, see Table \ref{Table:photometryfar}). The best graybody fits yielding the indicated values for T$_D$ and $\beta$ are shown. Fluxes at $\lambda$ $\leq$ 2.0 mm (log $\nu$ [GHz] $\geq$ 2.1) are weighted stronger in the fits.}
\label{Fig:SEDslong}
\end{figure*}

\end{document}